\documentclass[prb,aps,showpacs,twocolumn,superscriptaddress,10pt,amsmath,amssymb]{revtex4-1}
\pdfoutput=1
\usepackage{amsmath}
\usepackage{amssymb}
\usepackage{graphicx}

\newcommand{\bise}{Bi$_2$Se$_3$}
\newcommand{\bite}{Bi$_2$Te$_3$}
\newcommand{\sbte}{Sb$_2$Te$_3$}
\newcommand{\sbse}{Sb$_2$Se$_3$}
\newcommand*{\bb}[1]{\boldsymbol{#1}}
\newcommand*{\Zd}{$\mathbb{Z}_2$}

\makeatletter 

\begin{document}

\title{\it  Strain Effects in Topological Insulators: Topological Order and the Emergence of Switchable Topological Interface States in \sbte/\bite\ heterojunctions} 

\author{H. Aramberri}
\affiliation{Instituto de Ciencia de Materiales de Madrid, ICMM-CSIC, Cantoblanco, 28049 Madrid, Spain.}

\author{M.C. Mu\~noz}
\affiliation{Instituto de Ciencia de Materiales de Madrid, ICMM-CSIC, Cantoblanco, 28049 Madrid, Spain.}

\date{\today}
\begin{abstract}
We investigate the effects of strain on the topological order of the 
\bise\ family of topological insulators by {\it ab-initio} first-principles 
methods. Strain can induce a topological phase transition and
 we present the phase diagram for the 3D topological insulators, \bite, \sbte, \bise\ 
and \sbse, under combined uniaxial and
 biaxial strain. Their phase diagram is universal and shows metallic and insulating phases, both topologically 
trivial and non--trivial. In particular, uniaxial tension can drive 
the four compounds into 
a topologically trivial insulating phase. We propose a \sbte/\bite\ heterojunction
 in which a strain-induced topological interface state arises in the common gap of this 
normal insulator--topological insulator heterojunction.
Unexpectedly, the interface state is confined in the topologically trivial
 subsystem and is physically protected from ambient impurities. 
It can be switched on or off by means of uniaxial strain and therefore 
\sbte/\bite\ heterojunctions provide a topological system which hosts tunable
robust helical
 interface states with promising spintronic applications.
\end{abstract}

\maketitle

\section{Introduction and Motivation}
\label{sec:intro}
 Topological insulators (TIs) are a novel
 quantum phase of matter characterized by a topological invariant~\cite{FuKaneMele,qshe3d,FuKane} that
 exhibit topologically protected states at the boundary with a trivial insulator~\cite{hasan:2010}.
 In particular, the \bise\ family of three--dimensional (3D) TIs  
 has been extensively studied during the last few years as paradigmatic TIs that show an inverted
 band gap due to a strong spin--orbit coupling (SOC)~\cite{zhang2009topological}. At the surface,
 these materials exhibit a Dirac cone--like
 helical state with a circular skyrmionic spin texture~\cite{hexwarpspinz}, and the topological protection
 ensures the robustness of these states against disorder scattering as long as time--reversal symmetry is maintained.

 Fundamental interest and potential applications have driven the search of 
 external and internal agents such as stress, electromagnetic fields, chemical
 substitution or stacking defects~\cite{tunabledirac}, to engineer and manipulate the band structure
 of TIs. In particular, strain can be exploited to control the topological order.
 Several works have already assessed the importance of purely uniaxial
 strain in these materials and its influence on their topological
 character~\cite{liu2014tuning,bisestress2,bisetestresses,abinitiostress,stresssbtebisete,sbsebisestress,aplstress}. 
 For bulk materials, it was predicted that the 
 topological phase can be effectively manipulated by strain~\cite{bisestress2}.
Uniaxial strain can be induced by
 the chemical intercalation of zerovalent non--magnetic metals in the van der Waals (vdW) gaps.
 This technique has already been experimentally demonstrated and developed 
by Koski \textit{et al.}~\cite{zerovalent} in \bise\ to effectively enhance
the $c$ lattice parameter without disrupting the ionic or electronic 
configuration. In addition, \bise\ films under tensile stress along the 
$c$-axis have been recently grown via a self-organized order method and 
significant changes of the Fermi level and band gap of those films  
have been measured~\cite{Kim2016}. Topological state shifts at
 the strained grain boundaries in \bise\ films have also been reported~\cite{liu2014tuning}.
 To our knowledge, no study 
 has systematically addressed the combined effect of both uniaxial
 and biaxial strain in the topology of the \bise\ family.  
 Being the four compounds narrow gap semiconductors, small strain fields
 can strongly affect their electronic properties, and, consequently,
 their topological nature.
 In this work we study the role of combined uniaxial and biaxial
 tension on the \bise\ family of compounds, namely \bite, \sbte, \bise\ 
 and \sbse. 

 We show how uniaxial and biaxial strain can tune several properties of the 
 topological states and 
 how the combined effect of both kinds of strain can drive 
 the four systems into a metallic phase or two topologically distinct insulating phases. 
 We calculate the phase diagram for the four  
 materials in terms of uniaxial and biaxial strain, and we show the band inversion process that governs their topology.
 Furthermore, we predict the emergence of strain induced topological
 interface states in \sbte/\bite\ heterojunctions. The article is structured as 
 follows: in Section~\ref{sec:model} we describe the methods 
 employed for the calculations along with the crystal structure of the \bise\ family of compounds.
 Section~\ref{sec:strain} is devoted to the effect of uniaxial and biaxial 
 strain in bulk and thin films of the studied compounds.
 Next, we propose two topologically distinct heterojunctions of \bite\ and \sbte\ and 
 address their special electronic properties in Section~\ref{sec:heterojunctions}. Finally, Section~\ref{sec:summ} includes
 a summary of the results and conclusions.

\section{Methods and Crystal Structure}
\label{sec:model}
\begin{figure}
 \centering
\includegraphics[scale=0.230]{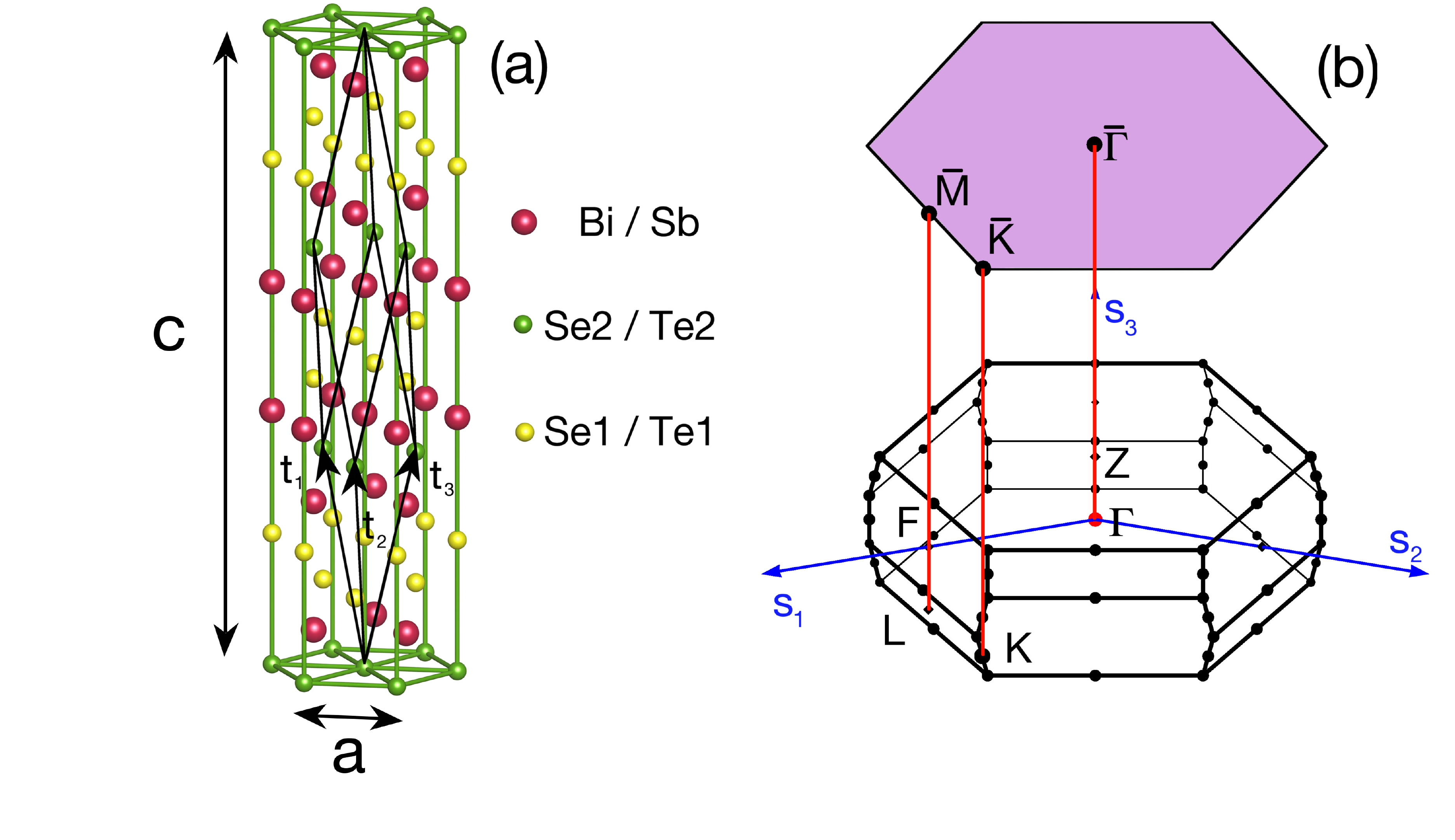}
 \caption{(a) Rhombohedral unit cell of the four studied compounds.
 Lattice constants $a$ and $c$ are indicated in the figure. (b) The
 corresponding bulk Brillouin Zone along with its projection along the
 [111] direction (purple shaded area).
}
 \label{geomBZ}
\end{figure}

 Bismuth dichalcogenides show a rhombohedral
 crystal structure with a five atom basis that constitute
 a quintuple layer (QL) --see Figure~\ref{geomBZ}\mbox{--.} The four
 compounds forming the \bise\ family studied in this work belong
 to the $R\overline{3}m$ ($D^5_{3d}$) crystallographic group.
 Along the [111] direction
 each atomic layer contains only one element and
 is hexagonally compact. The stacking pattern along this direction
 is ...AbCaB... where capital (small) letters indicate the position
 of Se or Te (Bi or Sb) atoms. Within a QL,
 interactions among the atoms are strong, while inter--QL bonding is of the weaker 
 vdW kind.

 To model the systems we employed the Vienna \textit{ab-initio} simulation
 package (VASP)~\cite{VASP-Kresse2-PhysRevB.48.13115}
 density functional theory (DFT) code for the atomic relaxations and electronic structure calculations of bulk materials.
 The SIESTA code~\cite{soler2002siesta}, through its implementation in the GREEN package~\cite{green},
 was additionally used for electronic structure calculations
 of the \bite--\sbte\ heterojunctions. In all the calculations we used the Perdew--Burke--Ernzerhof~\cite{pbegga}
 implementation of the generalized gradient approximation (GGA). The semi--empirical
 pair--potential vdW correction
 of Grimme~\cite{vdW-VASP_GrimmeDFT-D2} was used in the atomic relaxations as implemented in the VASP
 code to correctly account for the weak inter-QL interaction.
 The spin--orbit coupling was included self--consistently in both VASP~\cite{vaspsoc} and
 SIESTA--GREEN~\cite{greensocjorge} calculations. A 340~eV energy cut--off was employed for the plane wave basis 
 set in VASP calculations, while a double $\zeta$-polarized scheme with confinement 
 energies of 100~meV was used for the numerical atomic orbital basis set in SIESTA.
 Three--center integrals in SIESTA were computed using an hyperfine mesh cut--off 
 of 1200~Ry, equivalent to a real space grid resolution below 0.05~\AA$^3$.
 Biaxial (uniaxial) strain was taken into account by elongating or contracting lattice parameter
 $a$ ($c$) --see Figure~\ref{geomBZ}-- and allowing the internal coordinates of the ions to 
 relax. Biaxial and uniaxial strain ($\epsilon_a$ and $\epsilon_c$ respectively) of 
 a particular compound with lattice parameters $a,c$ are given by:
 \begin{equation}
 \begin{aligned}
\epsilon_a&=(a-a_{eq})/a_{eq} \\
\epsilon_c&=(c-c_{eq})/c_{eq}
 \end{aligned}
 \end{equation}
 where $a_{eq}$ and $c_{eq}$ are the equilibrium values of the in--plane and out--of--plane lattice parameters
 respectively.

\begin{figure}
 \centering
\includegraphics[scale=0.25]{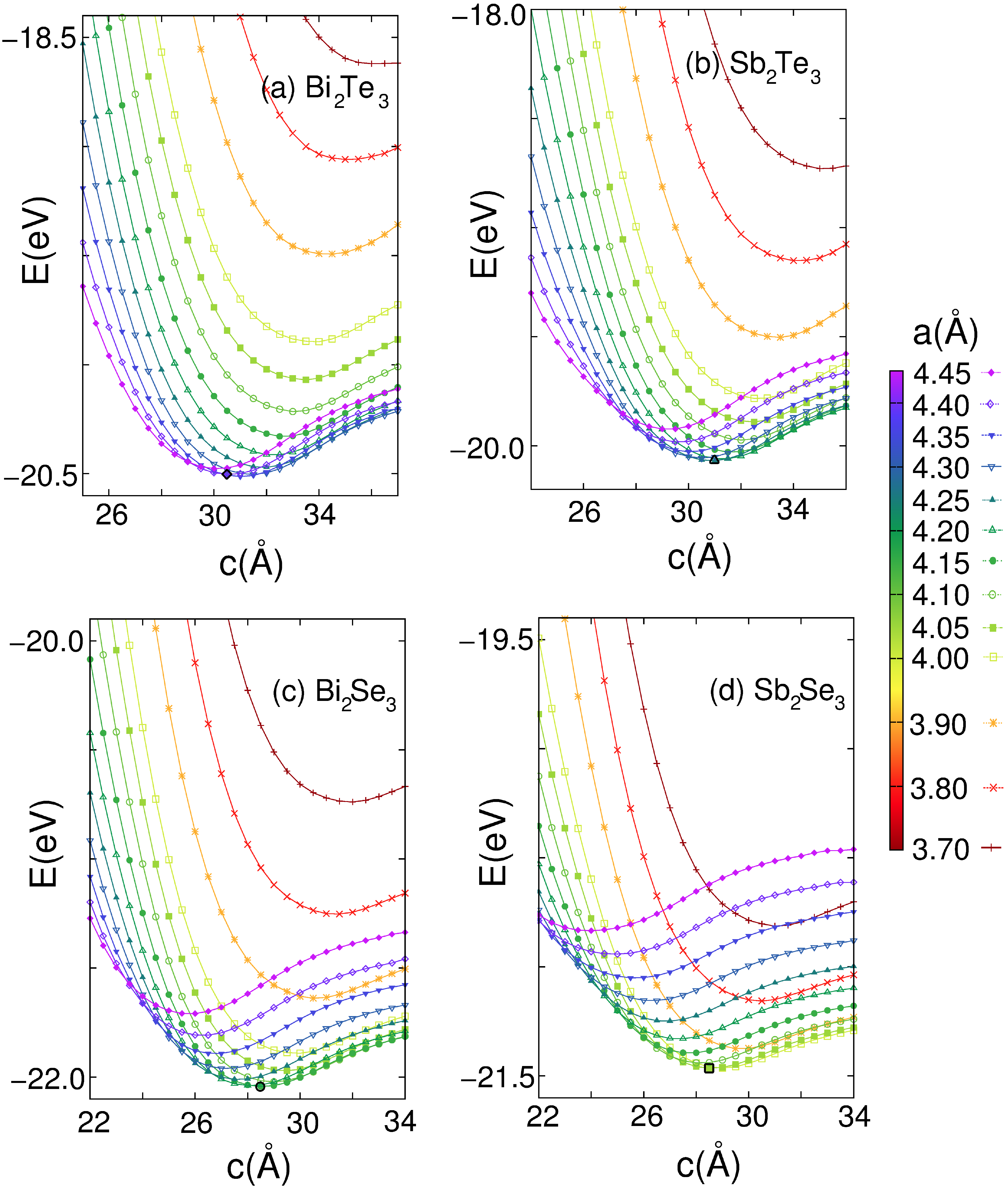}
 \caption{Total energy calculations using GGA+vdW in VASP for
 (a) \bite, (b) \sbte, (c) \bise\ and (d) \sbse. Each curve  shows
 the energy versus $c$ for a fixed value of $a$ (see legend at the right).
 Symbols marked in black indicate the equilibrium configuration. The exact values of 
 the equilibrium lattice parameters are given in Table~\ref{latparam}.
}
 \label{vdwlattice}
\end{figure}

\section{Uniaxial and Biaxial Tension}
\label{sec:strain}
\subsection{Bulk Materials}
\begin{figure}
 \centering
\includegraphics[scale=0.50]{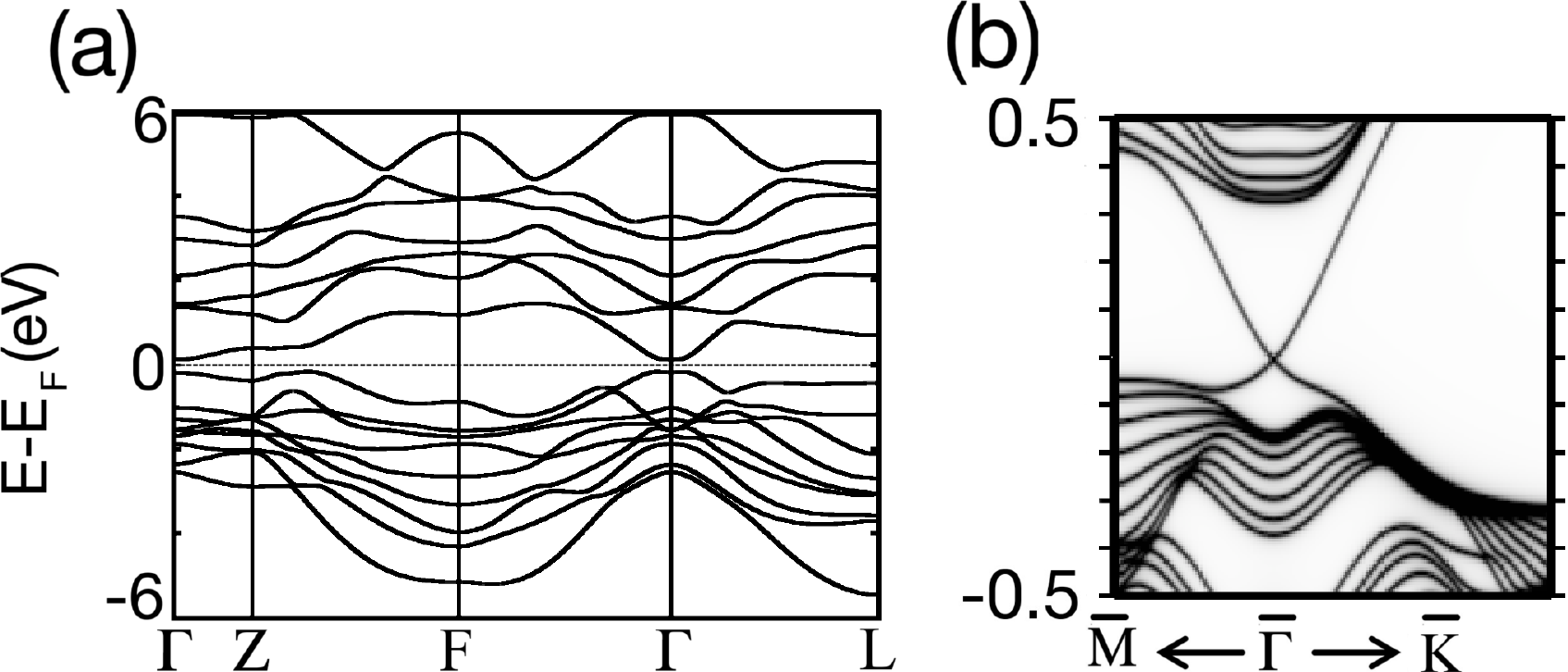}
 \caption{(a) Bulk band structure of fully relaxed \bise. The system is insulating and 
shows an inverted band gap. (b) Band dispersion of a 6~QL \bise\ slab. Topological surface states
 with a Dirac--like dispersion expand the bulk band gap, evidencing the topological character
 of unstrained \bise.
}
 \label{unstrainedbise}
\end{figure}

To address the effects of biaxial tension, we first calculated
the total energy of the \bise\ family of compounds for different values of the
lattice constants using GGA+vdW with the VASP code. In this way,
we obtain the relaxed geometry for a fixed value of the in--plane
lattice parameter $a$. Figure~\ref{vdwlattice}
shows the total energy of \bite, \sbte, \bise\ and \sbse\ for different
values of $a$ as a function of the out--of--plane lattice constant $c$.
 The equilibrium lattice parameters $a_{eq}$ and $c_{eq}$ were also 
calculated, and are given in Table~\ref{latparam}.
 In Figure~\ref{unstrainedbise} we show the band structure of fully relaxed \bise\ 
 for both bulk and thin film geometries as a reference. As shown 
in Figure~\ref{vdwlattice},
 for compressive in--plane biaxial strains ($\epsilon_a<0$)
 lattice parameter $c$ tends to increase, while for tensile strains
 ($\epsilon_a>0$)  $c$ decreases with respect to 
its equilibrium value. In fact, we can estimate the value 
 of the Poisson ratio $\nu$ from our calculations with
  the following equation for equibiaxial strained systems~\cite{ohring}:
 \begin{equation}
 \label{poisson}
\epsilon_c=-\frac{2\nu}{1-\nu}\epsilon_a
 \end{equation}

\begin{table}
 \centering
  \begin{tabular}{ccccc}
               & \bite          & \sbte         & \bise         &  \sbse      \\ \hline
$a_{eq}$ (\AA) & 4.40  (4.383)  & 4.25 (4.25)   & 4.17  (4.138) &  4.04  (-)   \\
$c_{eq}$ (\AA) & 30.5  (30.487) & 30.9 (30.35)  & 28.4  (28.64) &  28.7  (-)    \\
\end{tabular}
\caption{Calculated values of the equilibrium lattice parameters of the \bise\ family of compounds.
  Relaxations were carried out with VASP in the GGA+vdW approximation.
  Values in parentheses correspond to experimental data from Ref.~[\onlinecite{wyckoff}].
  No experimental data was available for \sbse\ in the rhombohedral phase.}
 \label{latparam}
\end{table}
  Fitting the energy minima positions for the curves shown in Figure~\ref{vdwlattice}
  to eq.~\ref{poisson}
  we obtain Poisson ratios of 
  0.30, 0.32, 0.29 and 0.27 for \bite, \sbte, \bise\ and \sbse\ respectively, in agreement
  with previous calculations~\cite{sbsebisestress,gao2016first,koc2013structural}.

\begin{figure*}
 \centering
\includegraphics[scale=0.70]{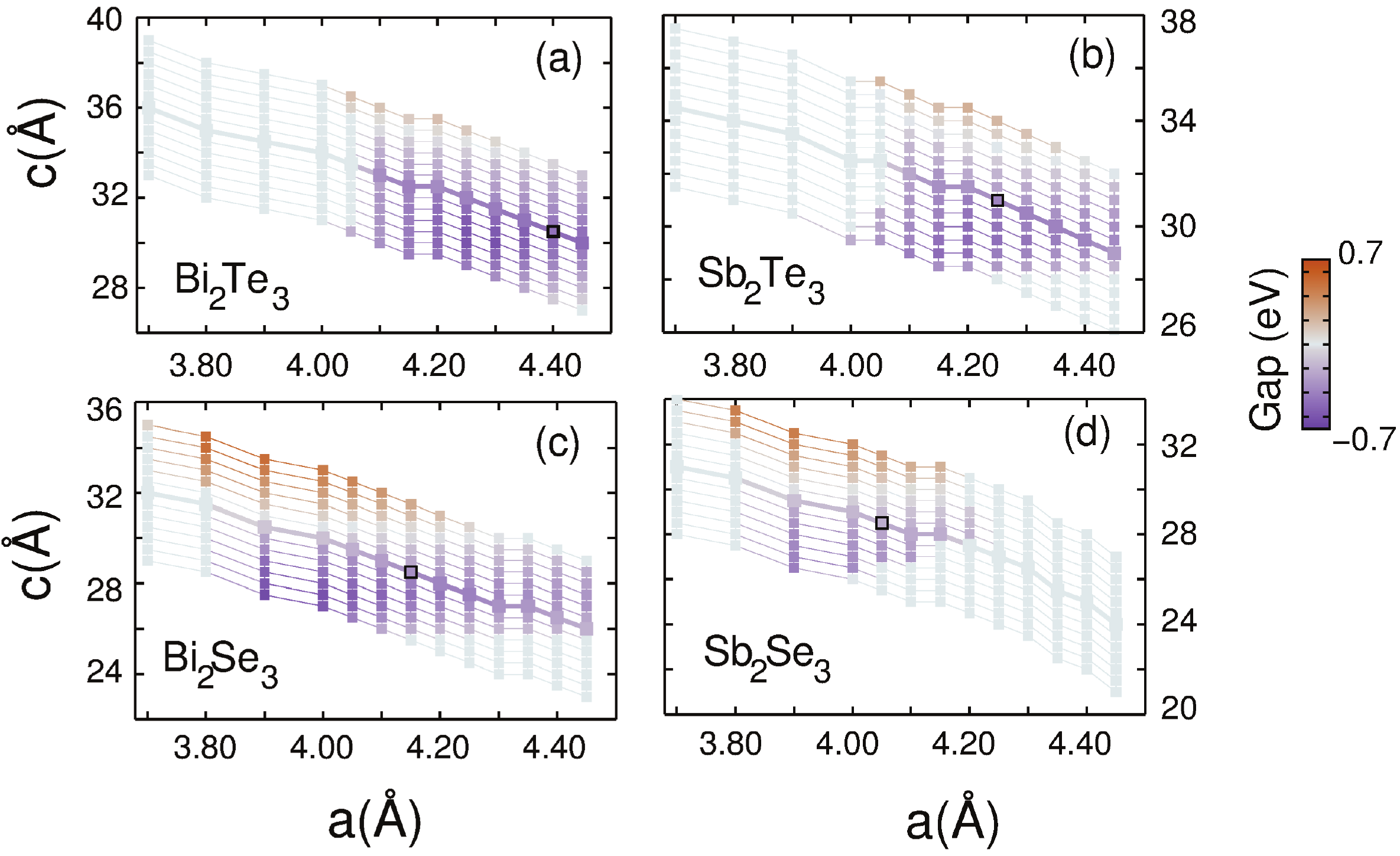}
 \caption{(a) Band gap (in color code shown on the right) for \bite\
for different values of lattice constants $a$ and $c$. 
The ionic configurations were allowed to relax for every single calculation.
Negative gaps (purple) indicate that the energy of Bi $p_z$ orbital
is lower than the Se $p_z$ orbital, i.e., the system is topologically non--trivial.
(b), (c) and (d) show equivalent diagrams for \sbte, \bise\ and \sbse\ respectively. 
These  phase diagrams show the regions in phase space
where the system is a normal insulator (orange),
a TI (purple) or a metal (gray). The equilibrium
 position is marked with a black square and the thicker line corresponds to the
relaxed $c$ lattice parameter for a fixed value of $a$ 
for each system.
}
 \label{phasediagram}
\end{figure*}

\begin{figure}
 \centering
\includegraphics[scale=0.50]{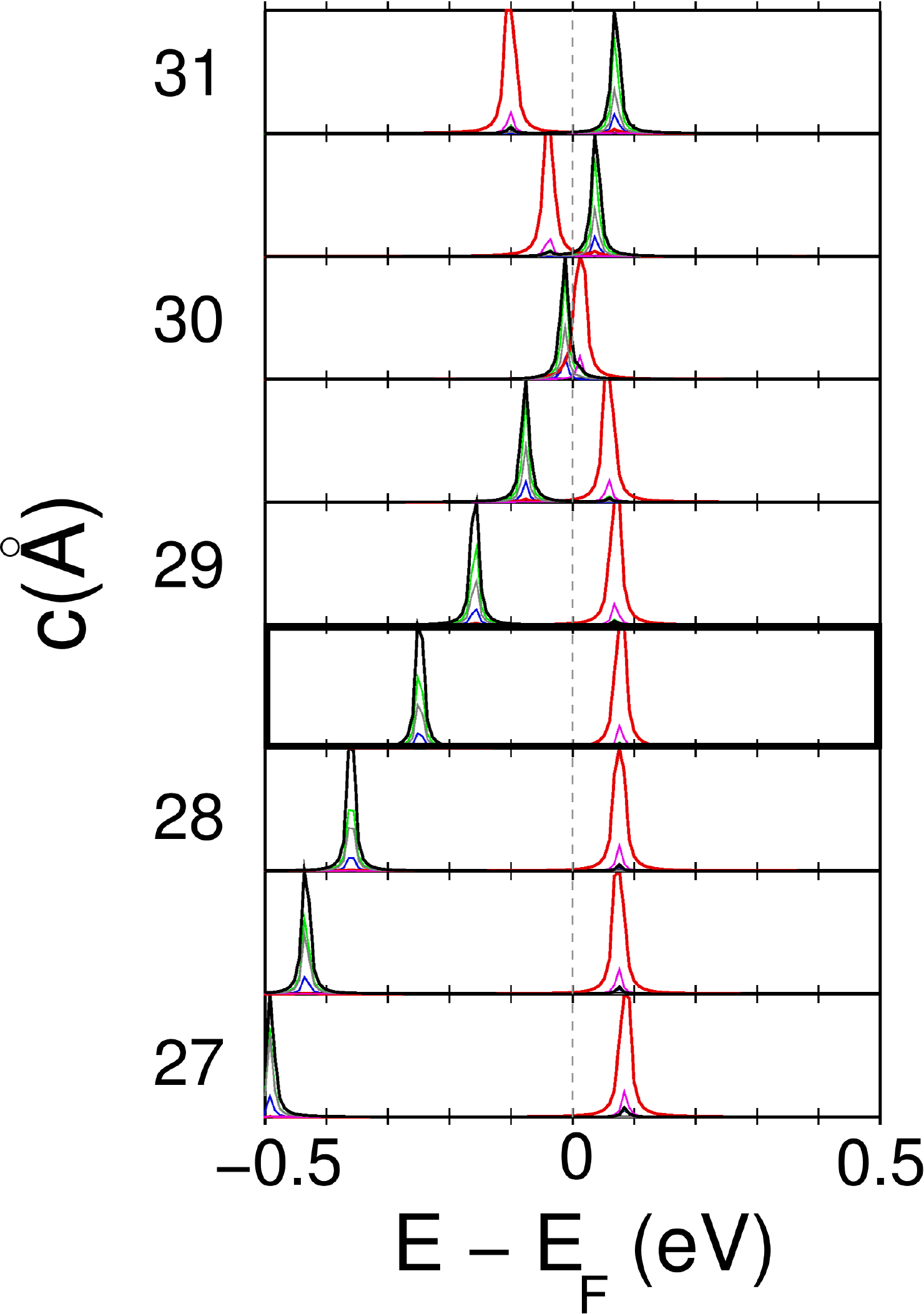}
 \caption{Projected density of states (PDOS) in the $\Gamma$ point around the energy gap
for bulk \bise\ with lattice parameter $a$ fixed to 4.20~\AA\ and 
different values of $c$ (left axis). The equilibrium configuration
is marked with a thicker frame. Black and red lines show
the contribution of Bi $p_z$ and Se $p_z$ orbitals respectively.
The magenta and blue lines indicate the Bi and Se $s$ contributions,
while the gray and green lines depict the Bi and Se $p_x+p_y$
weight. At $c\sim$30~\AA\ the bulk band gap closes and the system undergoes
a topological phase transition, so that for $c>(<)$~30~\AA\ the bands
are \textit{uninverted} (inverted) and the system is topologically
trivial (non--trivial). The different behavior of the Se and Bi $p_z$ bands with 
uniaxial strain is apparent in the figure.
}
 \label{dosbulk}
\end{figure}

In order to address the combined effects of out--of--plane uniaxial
 and in--plane biaxial strain,
 the phase diagram of the \bise\ family of compounds was computed
for several points in parameter space $(a,c)$. For every pair of values
of the lattice constants, we allowed the ionic positions to relax
and we calculated the energy spectrum. In this way we can address the
combined effect of uniaxial (along the $c$ direction)
and biaxial strain.
The results are summarized in Figure~\ref{phasediagram} for the four compounds.
For a fixed value of $a$, points 
along vertical lines correspond to increments of $\sim$ 1.5\% of uniaxial strain,
being the central point of each vertical line the relaxed value of $c$ for
the given amount of biaxial strain in that line.
For the four systems, three distinct phases can be identified:
a metallic phase, a topologically trivial insulating phase --normal insulator (NI), with 
\Zd\ topological invariant 0 -- and 
a topologically non--trivial phase --topological insulator (TI), with \Zd\ topological invariant 1 --.
The metallic phase is obtained for large in--plane biaxial strains in any direction. This is
due to the fact that a high compressive in--plane strain enhances the
bandwidth of the $p_x$ and $p_y$ orbitals in the valence band (VB), which eventually 
crosses the Fermi level and makes the system metallic. For high tensile
in--plane biaxial strains, the conduction band (CB) undergoes an analogous process, leading 
also to a metallic system. 
For moderate in--plain strains (below $\sim$~10\% in absolute value) the systems
remain insulating. In this range, the topological behavior of these
systems is governed by the band inversion between the Se and Bi $p_z$ bands,
and a topological phase transition (TPT) can be induced by out--of--plane strain.
 Starting from an inverted phase, for
 $\epsilon_c<0$ the bandwidth of the $p_z$ bands is enhanced,
which in turn makes the gap bigger at first, until eventually the gap becomes indirect, starts
to decrease and at a certain large compressive out--of--plane strain the
 system becomes metallic again. On the other hand, tensile out--of--plane
 strain ($\epsilon_c>0$) tends to diminish the gap until it closes when the energies of the
 Bi and Se $p_z$ bands at the $\bar{\Gamma}$ point become equal.
Further tensile strain reopens the gap, turning the system into
a topologically trivial insulator.
 Figure~\ref{dosbulk} shows the behavior
of the gap with out--of--plane strain for \bise\ at $a$=4.20~\AA.
 The band gap closing
and reopening is evident from the crossing between the Bi and Se $p_z$ bands,
 which have opposite parity and are responsible for the topological nature of the
 \bise\ family of compounds~\cite{zhang2009topological}.

For $\epsilon_a$=0, the critical uniaxial strain driving the 
TPT for \sbse, \bise\, \sbte\ and \bite\
is 3\%, 6\%, 6\% and 12\% respectively.
 This trend is in turn related to the crystal structure
\textit{and} the strength of the SOC in each system, being largest in \bite,
 smallest in \sbse\ and
intermediate in \bise\ and \sbte. Note that this 
values are given for zero biaxial in--plane strain, and the TPT will occur at different 
values of $\epsilon_c$ for $\epsilon_a\neq 0$ (see Figure~\ref{phasediagram}).
 Other studies have shown similar TPTs for \bise--like systems under purely
 uniaxial strain of
 6--10\%~\cite{liu2014tuning,bisestress2,bisetestresses,abinitiostress,stresssbtebisete,sbsebisestress,aplstress}, 
which is in good agreement with our results. 

Recent studies have revealed the importance of quasiparticle corrections~\cite{quasiparticleYazyev,GWAguilera}
and temperature effects~\cite{temperatureeffects}, which lead to a renormalization of the single-particle bands.
However, according to those studies the band inversion persists and the value of the
\Zd\ invariant remains unchanged. Therefore, the inclusion of both effects could 
slightly 
modify the values of the critical strains, but our results should remain
qualitatively correct. In fact, although DFT is known to underestimate band 
gaps, we find very good agreement between our computed band gaps and experimental data~\cite{wyckoff,black57}.

\begin{figure}
 \centering
\includegraphics[scale=0.60]{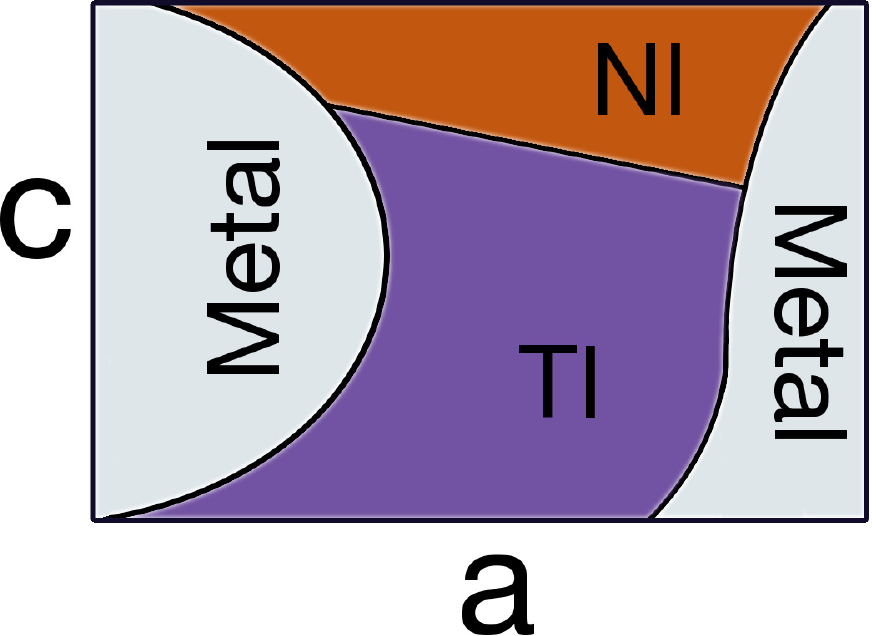}
 \caption{Schematic universal phase diagram for \bise--like systems
 in ($a,c$) parameter space. For high tensile and compressive in--plane
 strain the system becomes metallic. Uniaxial strain applied in the out--of--plane
 direction triggers a TPT.
}
 \label{phasepic}
\end{figure}

The universal phase diagram for the \bise\ family of 3D TIs under the combined effect of
 uniaxial and biaxial strain is sketched in Figure~\ref{phasepic}. 
To our best knowledge, no other previous work has systematically
addressed the 
effects on the topology of combined uniaxial and biaxial strain.
Moreover, as the four systems show a positive Poisson ratio, pure compressive biaxial strain 
induces an expansion in the $c$ direction which could, in principle, drive the system into
the normal insulating phase.
 Nevertheless, if no additional uniaxial strain is applied, we find that 
the four systems undergo a TI to metallic
phase transition with both tensile and compressive biaxial strain.
 The phase diagram 
 we provide for the four compounds can be useful for topological, band and orbital engineering of the \bise\ family of compounds in 
the fields of straintronics and spintronics.

 \begin{figure*}
 \centering
 \includegraphics[scale=0.50]{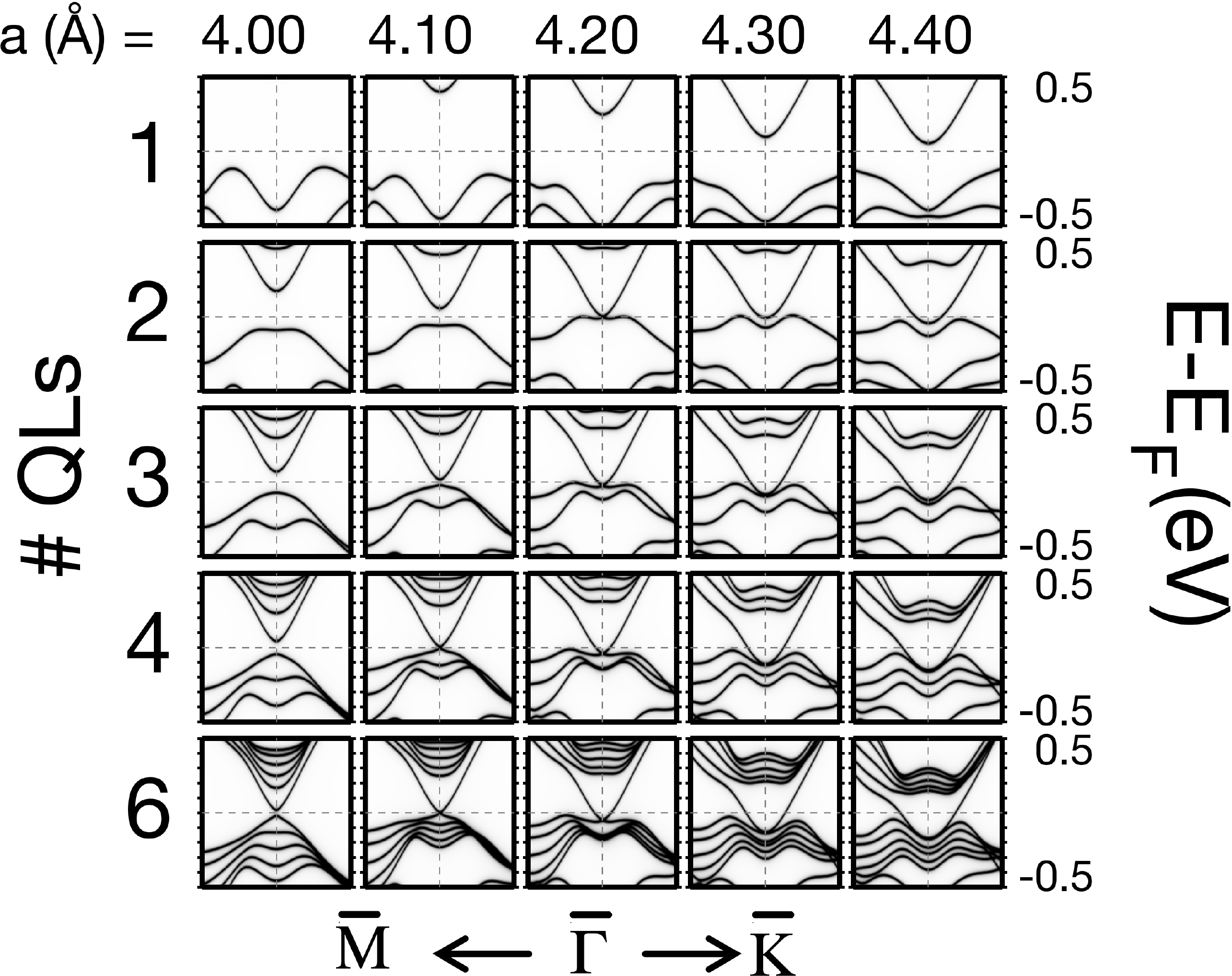}
 \caption{Band dispersion diagrams close to the $\bar{\Gamma}$ point along   
the $\bar{M}$--$\bar{\Gamma}$--$\bar{K}$ directions for \bise\ slabs under biaxial strain.
Each column corresponds to a fixed value of the in--plane lattice parameter labeled
on top and the corresponding out--of--plane $c$ parameter. The calculated equilibrium in--plane lattice parameter of unstrained 
\bise\ bulk is 4.17 \AA\ . The five rows correspond to different slab thicknesses: 1, 2, 3, 4 and 6~QLs
from top to bottom. The horizontal dashed lines indicate the Fermi level, whereas the 
vertical dashed lines show the $\bar{\Gamma}$ point.}
 \label{bisestress}
 \end{figure*}

\subsubsection*{The \sbse compound}
 Our results predict unstrained \sbse\ to be a topologically non--trivial insulator in the $R\overline{3}m$
  phase. Nevertheless, the region of parameter space in which \sbse\ is a TI is 
small and therefore minor variations of the lattice parameters result in a trivial insulator. Previous calculations
 have reported this material to be a normal insulator~\cite{zhang2009topological,sbsebisestress,sbsepressure}
 under no strain. Comparing with Refs.~\onlinecite{zhang2009topological,sbsepressure},
 we obtain slightly smaller lattice constants ($a$=4.04 versus 4.076~\AA, 
  $c$=28.7 versus 29.83~\AA\ for Ref.~\onlinecite{zhang2009topological},
  and similar values for Ref.~\onlinecite{sbsepressure}), probably due to the inclusion of 
  vdW corrections in our calculations. With their lattice constants
  our calculations also predict \sbse\ to be a narrow gap NI --see Figure~\ref{phasediagram} (d)--.
  In Ref.~\onlinecite{sbsebisestress} they obtain lattice parameters closer to ours ($a$=4.026~\AA, 
  $c$=28.732~\AA) within the GGA+vdW approximation, but decide to set 
  the equilibrium (unstrained) configuration at the plain GGA relaxed parameters
   (without the vdW correction, $a$=4.078~\AA, $c$=29.92~\AA), yielding again a NI phase.
  Recent calculations by other group~\cite{sbseti} estimate smaller values of lattice
  parameters for rhombohedral \sbse\ ($a$=4.004~\AA, $c$=28.553~\AA)
  and seem to predict an inverted band structure for antimony
  selenide --see the curvature of the bands around the $\Gamma$ point 
  in Figure 2 (d) of Ref.~\onlinecite{sbseti}--, but do not elaborate on its topological nature.
   Unfortunately, experimental data for \sbse\ is only available for its 
 more stable orthorhombic phase (\textit{Pnma})~\cite{sb2se3scirep}.
  We recently became aware of another work~\cite{sbseti2} in which DFT+vdW
   calculations predict rhombohedral \sbse\ to
    be topologically non--trivial.

\subsection{Thin Films} 
 Now we investigate the effect of pure biaxial in--plane strain along with low dimensional   
 effects on thin films of \bise--like systems. 
 Our starting points are bulk calculations in which, for a fixed
amount of biaxial strain, the lattice parameter $c$ was allowed to fully relax along with the
atomic coordinates. These bulks correspond to the equilibrium systems for each value of the in--plane strain and coincide with the minimum of each curve for fixed $a$ in Figure~\ref{vdwlattice},
 and with the thick lines in Figure~\ref{phasediagram}.
 Then, slabs of 1, 2, 3, 4 and 6~QL thicknesses were built
with the bulk positions and their band
dispersions were computed. 
 The results for \bise\ are shown in 
Figure~\ref{bisestress} 
for the range of lattice parameter $a$ in which the bulk system is a TI 
(see the results for \bite, \sbte\ and \sbse\ in the appendix, for the same range of lattice parameters).

The behavior with both compressive and tensile biaxial strain of \bise\ thin films is in clear analogy
 with the bulk behavior.
 Nevertheless, they show distinct features induced 
by strain. Under compressive biaxial strain 
 the size of the \bise\ bulk gap acquires a smaller value than that
of the unstrained system, and therefore the penetration depth of the surface states is enlarged
and a larger number of layers is needed to close the hybridization gap. Moreover, 
 the "M"--shaped feature in the VB around the $\bar{\Gamma}$ point is 
smoothed out and consequently the linear dispersion of the TSS is extended to a larger energy region in the VB.
 On the other hand, applying tensile biaxial strain also tends to close the bulk gap,   
 but the "M"--shaped feature becomes
 more pronounced and hence the DP in the films is shifted inside the VB 
(see the 6~QL series in Figure~\ref{bisestress}).
A sharp enough "M"--shaped VB detaches the DP from the Fermi level and consequently induces an $n$--type doping of the surface states. This result explains the shift in the
DP observed in Ref.~[\onlinecite{aplstress}], as well as the different behavior,
 gap opening or $n$-doping, observed at the grain boundaries in \bise\ films, in regions under compressive or 
tensile strain, respectively \cite{liu2014tuning}.

Moreover, a small decrease in the Fermi velocity with tensile strain is also apparent. 
The penetration depth of the TSS also varies with strain, and the closer in the phase diagram
to the critical Metal--TI lines the more QLs are needed to close the hybridization gap (see for instance the
 2~QL series in Figure~\ref{bisestress}, in which compressive or tensile biaxial strain takes the system
closer to a critical line in the phase diagram, and the TSS are gapped but for $a=4.20\approx a_{eq}$),
in agreement with the results displayed in 
Ref.~[\onlinecite{liu2014tuning}].
Higher compressive biaxial strain drives the thin films into a metallic state 
due to the upward shift in energy of the valence band maximum
(see for example Figure~\ref{bitestress} in the appendix), while for a critical
tensile biaxial strain the bulk--like CB crosses the Fermi level and
the thin films become metallic again (see Figure~\ref{sbsestress}).

 Strain can therefore turn the \bise\ family of compounds insulating or metallic, and allows for engineering
 of the gap, the orbital character of the bands, the Fermi velocity, DP energy
 and thus also the doping of the TSSs.
 The table--like figures for the four systems are displayed so that they can be used for
 determining what kind of band dispersion is expected when a bismuth dichalcogenide
of a certain thickness is grown on a substrate with a particular lattice parameter.

\section{Strained Heterojunctions}
\label{sec:heterojunctions}
When two distinct TIs are faced to one another, an interesting problem arises. If both
materials belong to the same $\mathbb{Z}_2$ topological class, no interface state
is guaranteed by the bulk--to--boundary correspondence, as the change in topological 
invariant is zero. Therefore, a topological surface state can be annihilated by placing
 another TI on top, even if both bulk gaps align in a 
straddling gap configuration. Still, topologically trivial interface states
may arise regardless of the topological invariants. 
 Moreover, in broken gap heterojunctions (without a common gap) no 
 topologically protected interface state may appear since the system will no longer be an insulator. 
In this section we will study interfaces of \bite\ and \sbte\, both in superlattices
 and in slab geometry. Among the four members of the \bise\ family of compounds, we have chosen these two
so that the difference in electronegativity, $\chi$, between the A and B elements in the A$_2$B$_3$
compounds is as small as possible, in order to obtain a straddling gap at the heterojunction
 and minimize the band bending along the system.
 Table~\ref{tab:electronegs} shows the Pauling and Allen 
electronegativities ($\chi_P$ and $\chi_A$) for Bi, Sb, Te and Se. The first two elements have an
 almost equal value
of the electronegativity --in fact $\chi_P($Bi$)<\chi_P($Sb$)$ while $\chi_A($Bi$)>\chi_A($Sb$)$--.
On the other hand, Se and Te show a bigger difference in their $\chi$ values. 
Opposite doping for Se- and Te-based materials is expected, and we have additionally calculated \bise/\bite\ heterojunctions
which exhibit a broken gap alignment, thus leading to a metallic phase where the $\mathbb{Z}_2$ invariant 
is ill-defined and no topological interface states can exist.

\begin{table}
 \centering
\begin{tabular}{ccccc}
                                    & Bi    & Sb     & Te     & Se    \\ \hline
  $\chi_P$~\cite{electronegPauling} & 2.02  & 2.05   & 2.1    & 2.55   \\
  $\chi_A$~\cite{electronegAllen}   & 2.01  & 1.984  & 2.158  & 2.434   \\
\end{tabular}
 \caption{Electronegativities of the four elements present in the \bise\ family of compounds according to
the Pauling scale (first row) and the Allen scale (second row). Bi and Sb present very 
similar values, whereas the difference between the electronegativities of Te and Se
is significant in both scales.}
\label{tab:electronegs}
\end{table}

\begin{figure}
 \centering
\includegraphics[scale=0.30]{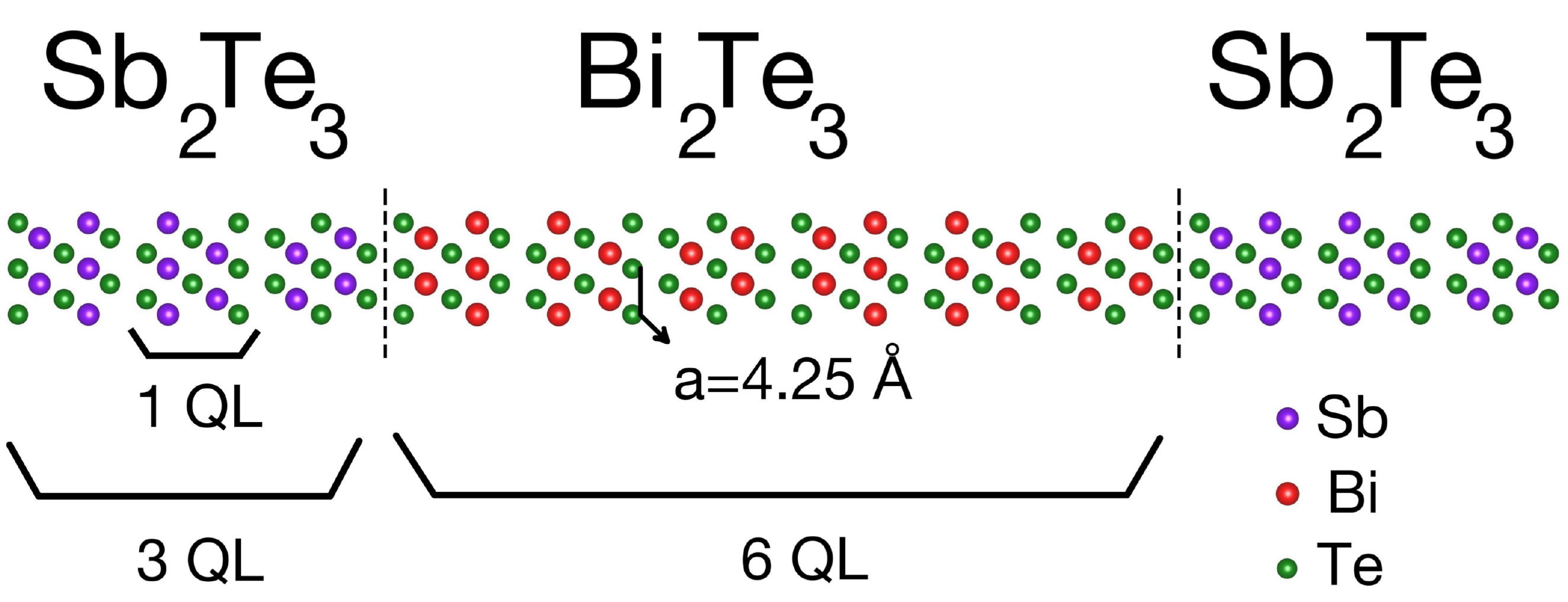}
 \caption{Geometry of the $m$-\sbte/$n$-\bite/$m$-\sbte\ trilayer with $m$=3 and $n$=6.
The in--plane lattice parameter $a$ is fixed to that of \sbte\ under no strain.
Interfaces are shown as dashed lines as a guide to the eye. The whole system follows the AbCaB stacking pattern
analogous to an \textit{fcc} (111) crystal, which ensures inversion symmetry is preserved.
 Superlattices are constructed by imposing periodic boundary conditions on this and similar trilayers,
which will also preserve inversion symmetry.
}
 \label{bisb363geom}
\end{figure}

\begin{figure*}
 \centering
\includegraphics[scale=0.26]{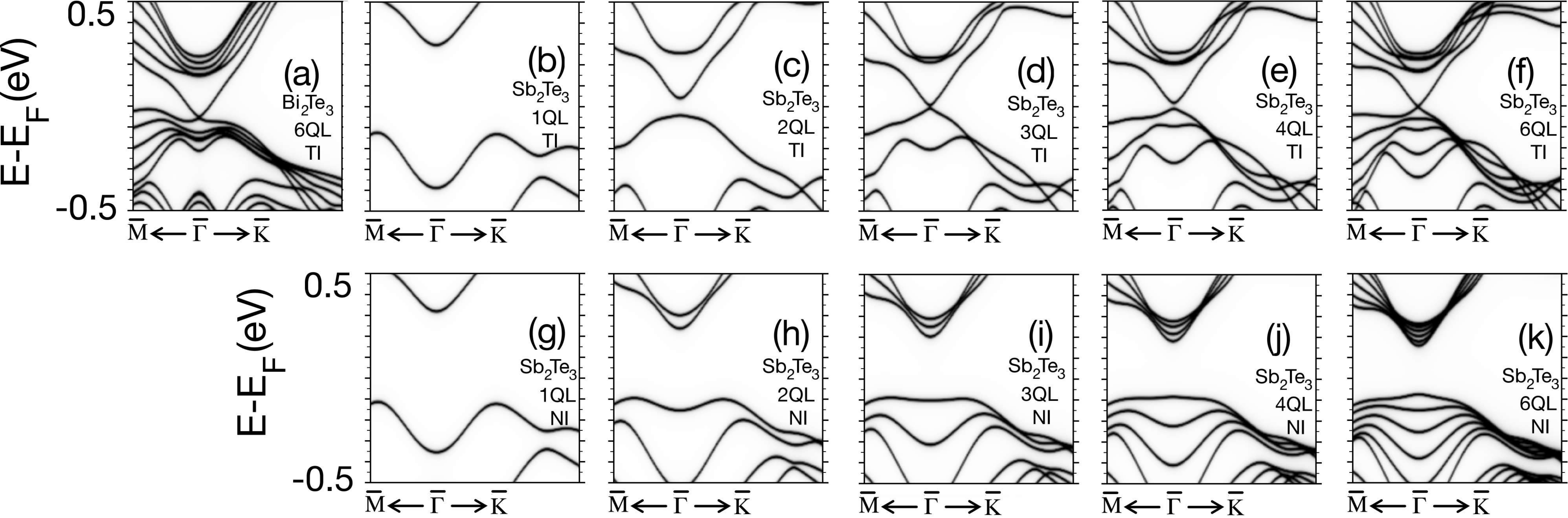}
 \caption{Band structure of the different isolated subsystems involved in the TI/TI/TI trilayers
 --top row, (a) to (f)-- and in the NI/TI/NI trilayers --bottom row and (a)--.
 (a) corresponds to a 6~QL \bite\ thin film under biaxial compressive strain so that its in--plane lattice parameter 
matches that of equilibrium \sbte\ (4.25~\AA). (b) to (f) show the band dispersion for unstrained 
 \sbte\ slabs of 1, 2, 3, 4 and 6~QLs respectively, which show a TSS according to their topologically non--trivial nature,
  although for thicknesses below $\sim$~5~QL a gap opens in the TSSs due to surface--surface interaction.
  (g) to (k) show the electronic structure of uniaxially elongated \sbte\ slabs of 1, 2, 3, 4 and 6~QLs respectively.
  In these cases the system is clearly in the normal insulating regime, since no TSS appears for thicknesses as large as 6~QL.
}
 \label{isol}
\end{figure*}

\subsection{TI/TI interfaces}
\label{ssec:TITI}

 We have chosen \sbte/\bite/\sbte\ trilayers with equal number of QLs of \sbte\ at both
 sides so that inversion symmetry is preserved, making the analysis simpler, as both interfaces
 will be equivalent. We calculated $m$-\sbte/$n$-\bite/$m$-\sbte\ trilayers, where $m$
 and $n$ are the number of QLs of \sbte\ and \bite\ respectively. 
In the superlattice geometry, due to periodic
 boundary conditions, the trilayer turns into a 2$m$-\sbte/$n$-\bite\ 
structure repeated in the [111] direction.
We still call it a $m$-\sbte/$n$-\bite/$m$-\sbte\ superlattice 
to emphasize the centrosymmetric nature of the system.
We fixed $n$=6, for which 
the surface--surface interaction in \bite\ is negligible and a gapless Dirac cone (DC) develops at the surface
 --see Fig.~\ref{isol}(a)--, while the number of \sbte\ QLs      at both sides is varied from $m$=1 to 3.
The AbCaB stacking sequence of the pristine subsystems is preserved along the interfaces and in 
the superlattices in order to preserve inversion symmetry.
The $C_3$ rotation axis and the three vertical mirror planes of the pristine systems are also preserved in the 
heterojunction.
We fix the in--plane lattice vector $a$ to that of \sbte\ in equilibrium, $a_{eq}$=4.25~\AA, 
and the $c$ lattice parameter for each subsystem is set to its relaxed value for $a$ fixed
to the aforementioned value, that is 30.9~\AA\ for \sbte\ and 32.0~\AA\ for \bite\
(see Fig.~\ref{vdwlattice}). The ionic coordinates within each subsystem are fixed to 
their relaxed bulk values, and the vdW gap between \sbte\ and \bite\ 
is taken as the average vdW gap between both subsystems. This setup could correspond to a 6~QL thick \bite\ slab
grown on a $m$--QL \sbte\ substrate, and another $m$--QL \sbte\ thin film grown on top of it.
 Figure~\ref{bisb363geom} depicts the geometry for the $m$=3 case. According to the phase diagram calculated in 
Fig.~\ref{phasediagram}, both the \sbte\ and \bite\ subsystems show an inverted gap in the bulk. This means that
the existence of an interface state is not guaranteed, since the change in the \Zd\ 
 invariant across the interface is zero as both materials are topological insulators.

\begin{figure}
 \centering
\includegraphics[scale=0.24]{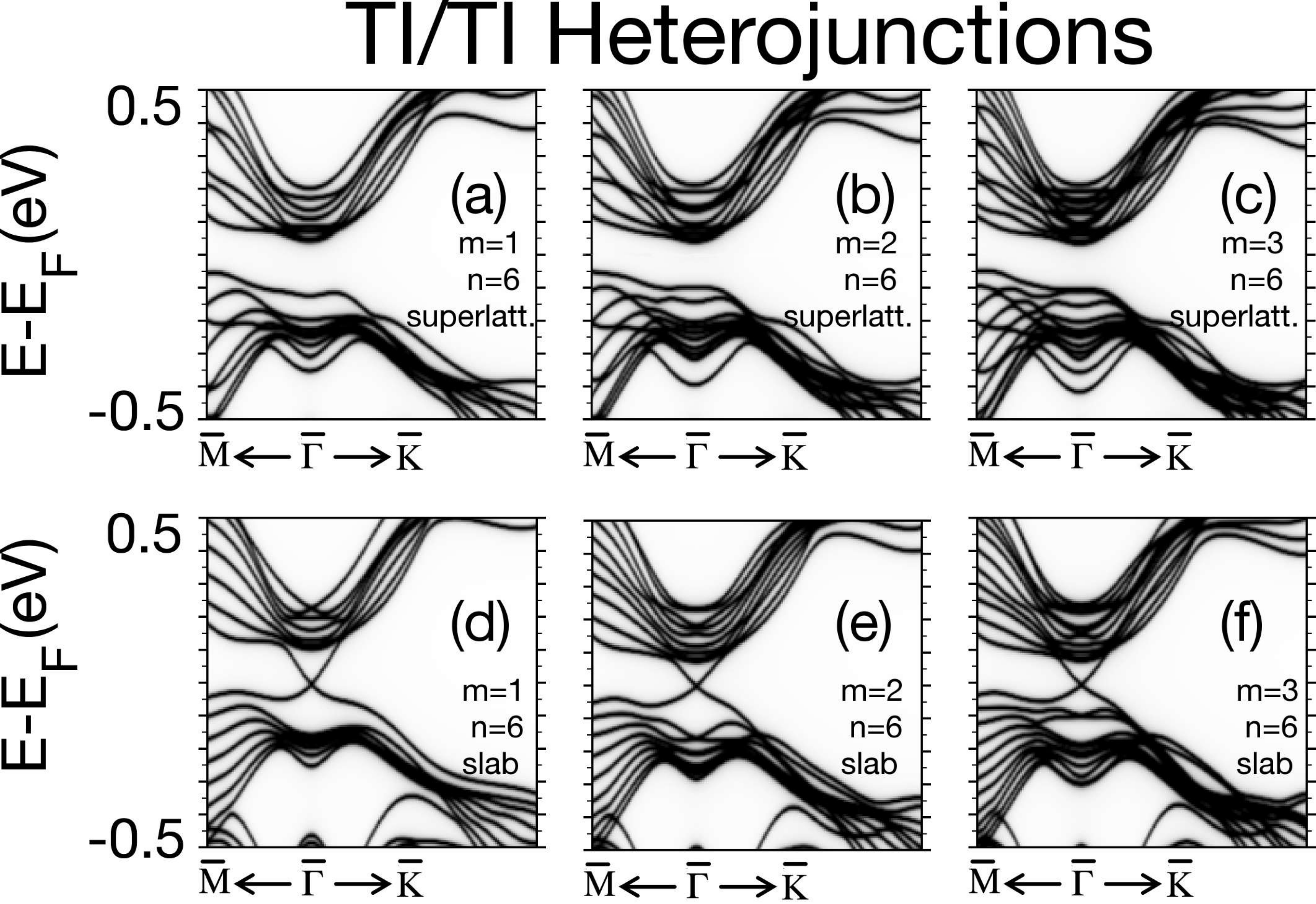}
 \caption{Band structure of the TI/TI/TI heterojunctions considered, 
 both in superlattice --(a) to (c)-- 
  and slab --(d) to (f)-- geometries.
(a), (b) and (c) show the band dispersion for the trilayers in a
 superlattice with $n$= 6 and $m$= 1, 2, and 3 respectively. As all the constituents of the superlattice
 are topologically non--trivial, there are no interfaces between subsystems with different value of the $\mathbb{Z}_2$ 
topological invariant, and no interface state exists.
 (d), (e) and (f)
 correspond to slab geometries with $n$= 6 and $m$= 1, 2, and 3 respectively. In these three cases
a surface state develops irrespective of the number of \sbte\ layers, but no interface state is present.
}
 \label{TI}
\end{figure}

\begin{figure}
 \centering
\includegraphics[scale=0.68]{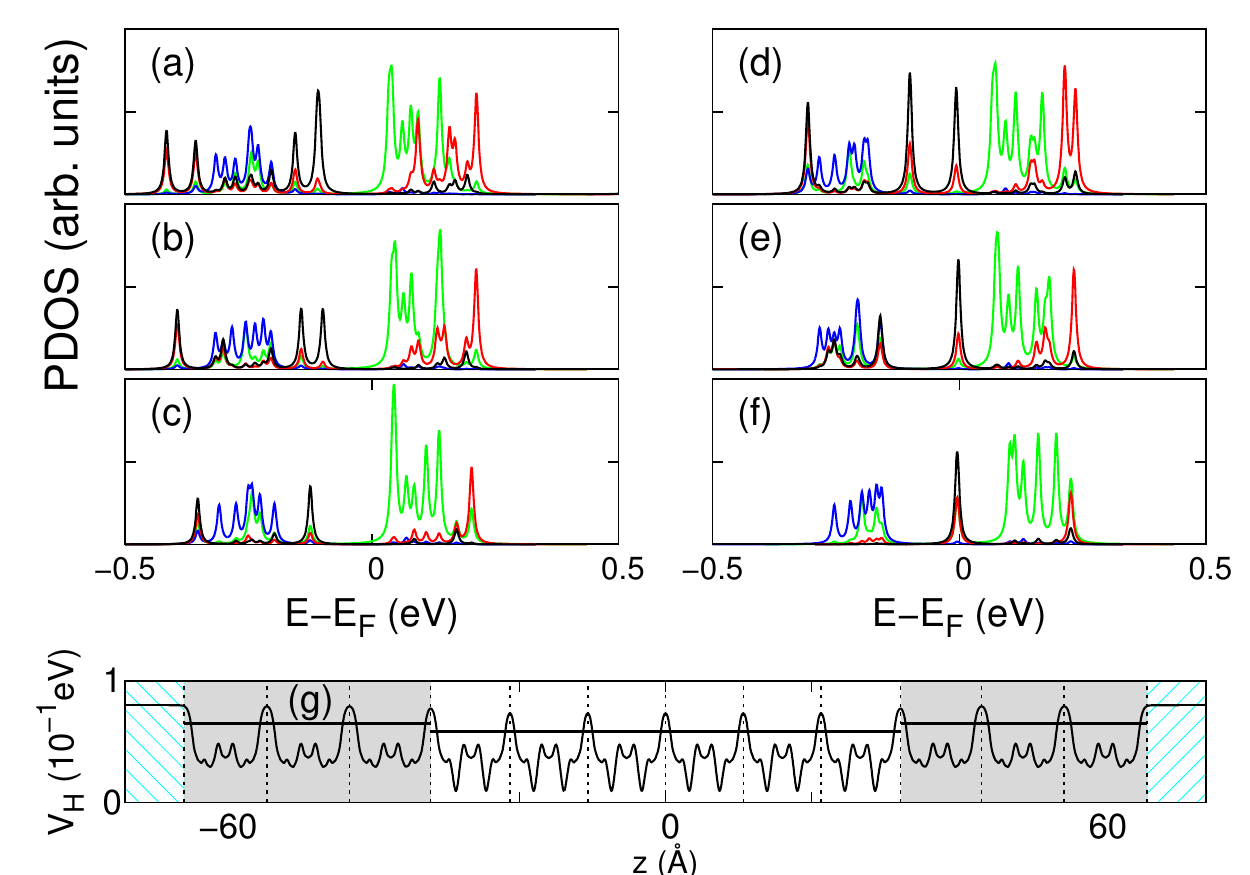}
 \caption{(a) to (f) show the PDOS in $\bar{\Gamma}$ close to the Fermi level of the studied TI/TI heterojunctions.
Blue and black lines show the contribution of Bi $p_z$ and Sb $p_z$ orbitals respectively.
The green and red lines indicate the Te $p_z$ contributions from the Te atoms in the \bite\ and \sbte\ subsystems respectively.
(a) to (c) --(d) to (f)-- depict the PDOS for superlattices --slabs-- with $m$=3 to 1. Both subsystems show an inverted band structure
 in all cases. The 2D averaged Hartree potential profile
 for the $m$=3 TI/TI slab is shown in (g), along with the average Hartree potential in each QL (horizontal straight solid lines).
}
 \label{dosTITI}
\end{figure}

\begin{figure}
 \centering
\includegraphics[scale=0.5]{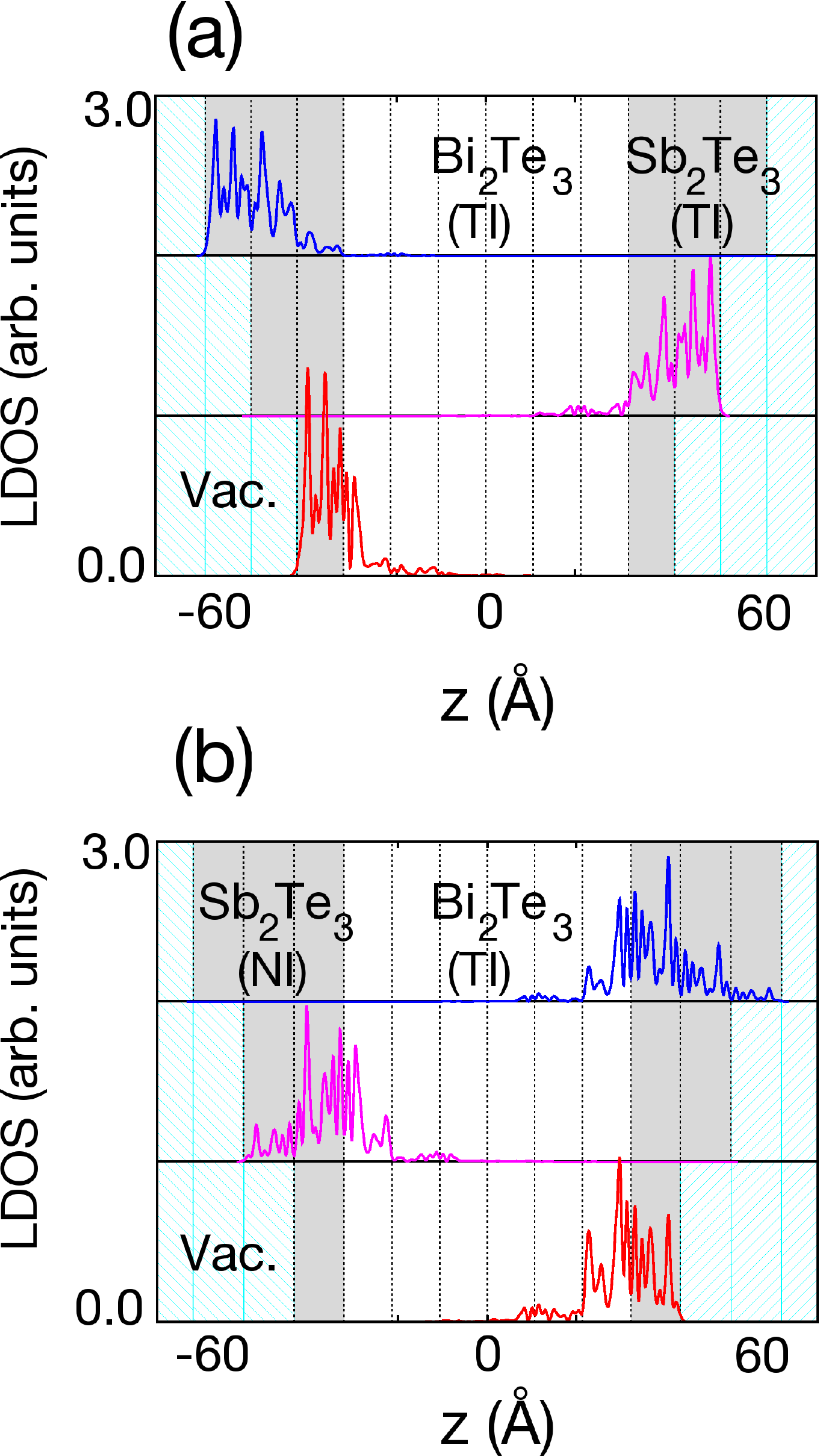}
 \caption{Layer projected density of states (LDOS) of the topological states in $m$-\sbte/$n$-\bite/$m$-\sbte\ 
trilayer slabs
for $m$=1 (red line at the bottom), $m$=2 (magenta line in the 
middle) and $m$=3 (blue line at the top).
The LDOS was computed at $\bb{k}$=(5, 0) $\cdot 10^{-3}$\AA$^{-1}$, along the $\bar{\Gamma}$--$\bar{K}$ direction
 and close to the $\bar{\Gamma}$ point for the electron--like TSSs.
The trilayer is centered at the middle of the 6~QL \bite\ layer. Vertical dashed lines
depict the boundaries of each QL. The gray shaded region corresponds to the \sbte\
 subsystem, while white regions belong to the \bite\ subsystem and the vacuum is shaded with a cyan pattern.
(a) shows the TI/TI/TI trilayer for which TSSs localize at the surfacemost QL of the \sbte\ which is in the
 topological insulating phase.
In (b) the LDOS of the NI/TI/NI trilayer is shown. In this case TISs localize at both interfacial QLs, one
 in the \bite\ subsystem and the other in the \sbte\ subsystem, being the latter in the NI phase. The extension
 of the TISs spans the whole trivial \sbte\ subsystem. 
 All the topological states shown are degenerate due to inversion symmetry,
 and only one of the two--fold degenerate states is shown in each case (being always the other state localized at the opposite
  surface or interface).
}
 \label{slabpens}
\end{figure}

We start by analyzing the electronic structure of the isolated subsystems, 
depicted in Figure~\ref{isol}. \bite\ under small
biaxial strain remains a TI, and so it develops surface states
 when truncated in the [111] direction. For a 6~QL slab
(as shown in the figure), surface--surface interaction is
 already negligible and the linearly dispersive DCs at the $\bar{\Gamma}$ point emerge. In contrast with
\bise\ or \sbte, the DP of the \bite\ surfaces is not at
 the Fermi level and lies below the VB maximum. 
This is in agreement with previous results~\cite{zhang2009topological},
 and can be attributed to the larger curvature of the 
VB along the $\bar{\Gamma}$--$\bar{M}$ direction.
\sbte\ in this system presents neither uniaxial nor biaxial strain,
 and it is therefore also in the topologically 
non--trivial phase. In Figure~\ref{isol} the electronic
 structure of unstrained 1, 2, 3, 4 and 6~QL thick \sbte\ films is 
also shown --panels (b) to (f)--. Antimony telluride presents a 
topological surface state being the DP at the Fermi level for 6~QLs.
 The penetration depth is $\sim$~2~QLs, so that a gap
opens in thin films of less than 5~QLs due to 
surface--surface hybridization.

The band structures of the periodic superlattices 
are shown in Figure~\ref{TI} (a) to (c) for $m$=1, 2 and 3, and those 
corresponding to the trilayer slabs in Figure~\ref{TI} (d) to (f).
The former --(a), (b) and (c)-- present a band gap of $\sim$~0.1~eV, being the VB (CB) offset
of $\sim$ 0.1 (0.05)~eV between both subsystems, with the VB (CB) of \sbte\ lying 
at a higher energy. 
The small band staggering at the heterojunction 
can be attributed to the small deviation in the 
values of the electronegativity for Bi and Sb. 
Figs.~\ref{dosTITI} (a), (b) and (c) represent the atomic orbital decomposed partial 
density of states (PDOS) at $\bar{\Gamma}$ in the energy region displayed in Fig.~\ref{TI}. 
They evidence the band inversion in both \sbte\ and \bite\ slabs and the 
similar band alignment for the three superlattices, $m$=1, 2 and 3. The top of the VB 
is dominated by Sb and Bi $p_z$ orbitals with positive parity, being the  
former at higher energy, while the Te $p_z$ orbitals with negative parity 
are located at the bottom of the CB region. The interaction between the CB Te 
orbitals of both compounds is weak, particularly for the wider superlattices. Hence,
in the superlattice both subsystems present an inverted band structure.

The bulk--to--boundary correspondence predicts no topologically 
protected interface state at the junction, and although trivial 
interface states could develop, our results show that this is not the case.  
We therefore conclude that this \sbte/\bite\ heterojunction 
is insulating with no interface states whatsoever, but will
 develop surface states when truncated.
This is proved in the thin film geometry --Figure~\ref{TI} (d),(e) and (f)--, where surface states that span 
the whole bulk band gap appear at both ends.
The DP of these TSSs is pinned at the Fermi level irrespective 
 of the thickness of the \sbte\ layers.
 In fact, even for the $m$=1 and 2 for which
the thickness of the \sbte\ subsystem is below the penetration
 depth of the surface states -see Figure~\ref{isol} (b) and (c)-, 
 there is no energy gap and the spectrum of the trilayer slab still 
exhibits a
semimetallic character. The corresponding atomic orbital PDOS are shown in 
Fig.~\ref{dosTITI} (d) to (f). The three trilayers exhibit a sharp peak at $\bar{\Gamma}$, 
associated with the TSSs. They have a predominant contribution of the \sbte\
orbitals, mostly of Sb $p_z$.

 This orbital contribution is consistent 
with the TSS localization shown in Figure~\ref{slabpens} (a), for the three different slabs.
The surface state is strongly confined in the \sbte\ subsystem, with a 
penetration depth of $\approx$ 2 QLs, although for
the $m$=1 case the state strongly localizes at the surface-most QL.
Fig.~\ref{dosTITI} (g) displays the 2D averaged Hartree potential profiles --including the 
ionic contribution-- along the [0001] direction for the $m$=3 slab.  
It reflects the chemical difference between both \sbte\ and  \bite\ 
compounds and the potentials are almost identical for the finite trilayers and the superlattice 
(not shown), differing only on the potential step at the surface of the slab.  
Therefore, our results corroborate the fact that unstrained \sbte\ is a TI 
even for ultrathin films, and support the idea that the TI/TI \sbte\--\bite\
heterojunction behaves as a homogeneous TI and does not confine neither topological 
nor trivial states at the interface.

\subsection{NI/TI interfaces}
\label{ssec:NITI}
Now we will discuss the effect of applying uniaxial 
tensile strain to \sbte\ in the 
system presented in the previous subsection. The systems considered 
are again $m$-\sbte/$n$-\bite/$m$-\sbte\ 
trilayers in either a slab geometry (thin film) or a superlattice.
The in--plane lattice parameter is again fixed to $a$=4.25~\AA\, and $c$ is set to 32.0~\AA\ for \bite, 
but now the \sbte\ subsystem is expanded to $c$=34.0~\AA,  
corresponding to an uniaxial tensile strain of -10\%.
According to the phase diagram shown in Figure~\ref{phasediagram},
 \sbte\ will now be in a normal
 insulating phase, so that at the interface of \sbte\ and \bite\ 
 the topological
\Zd\ invariant will increase from 0 to 1.  
 The TPT on \sbte\ can be induced by external uniaxial tensile strain 
or via 
 the chemical intercalation of zerovalent non--magnetic metals in the vdW gaps~\cite{zerovalent}
as stated in a previous section.

\begin{figure}
 \centering
\includegraphics[scale=0.24]{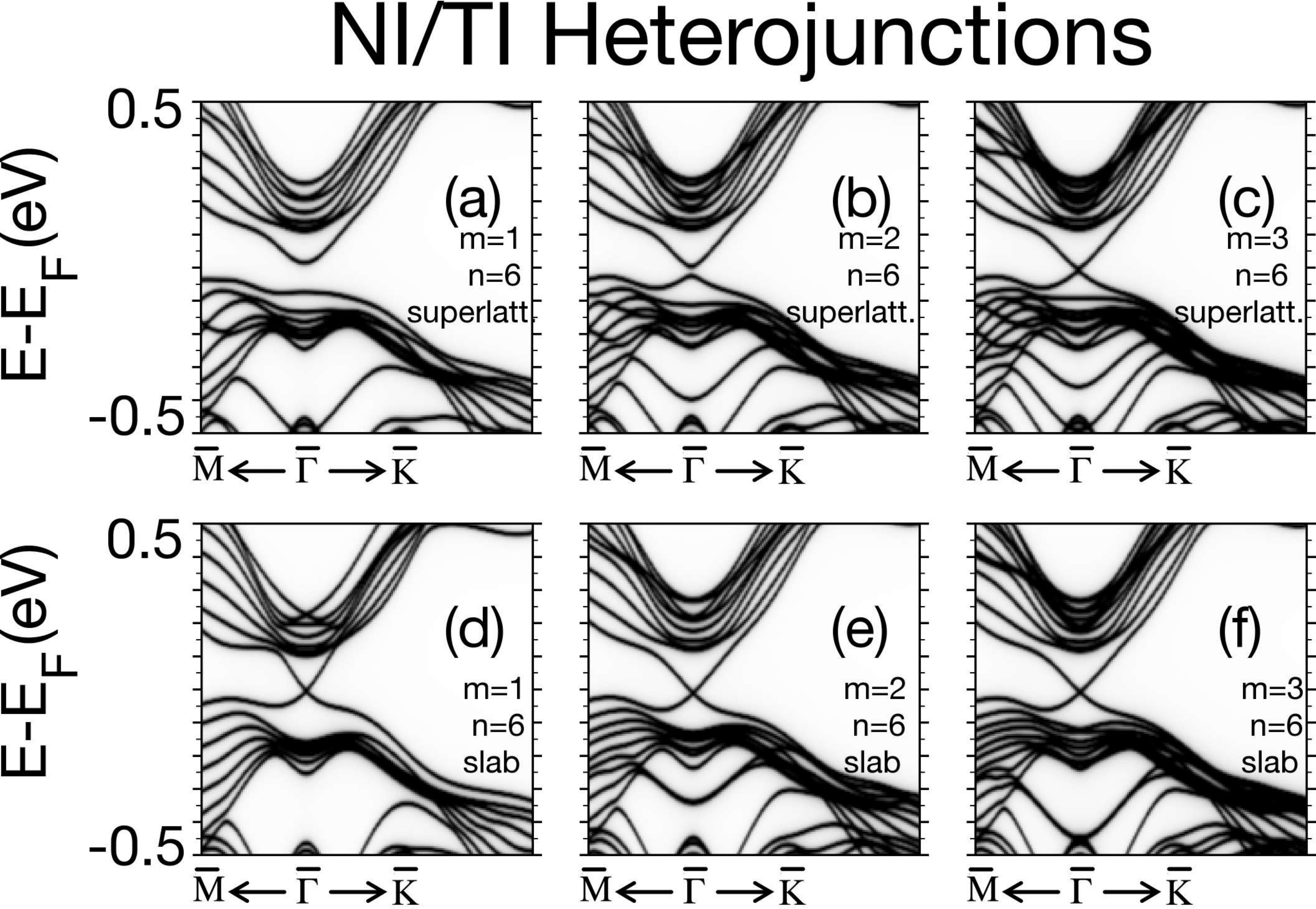}
 \caption{Band structure of the NI/TI/NI heterojunctions considered,
 both in superlattice --(a) to (c)-- 
  and slab --(d) to (f)-- geometries.
(a), (b) and (c) show the band dispersion for the superlattices 
with $n$=6 and $m$= 1, 2, and 3 respectively. As the \sbte\ subsystem has been driven to
the normal insulating phase by applying uniaxial tensile tension, topologically protected
states localized at the interface appear, according to the bulk--to--boundary correspondence.
A gap in the spectrum opens for $m$ below $\sim$~2
 due to interface--interface interaction (note that the total thickness of the \sbte\
subsystem is 2$m$ QLs in the superlattices).
 (d), (e) and (f)
 correspond to slab geometries with $n$=6 and $m$= 1, 2, and 3 respectively. In these three cases
a topological interface state (TIS) develops irrespective of the number of \sbte\ layers.
}
 \label{NI}
\end{figure}

The electronic structure of the different isolated constituents is shown
 in Figure~\ref{isol} (a) and (g) to (k). \sbte\ under such out--of--plane strain shows no
 band inversion in the bulk, and this is reflected in the thin film electronic
 structure. In contrast with the previously 
analyzed systems, \sbte\ now shows no surface state
 since it is in the NI phase. The gap is of 0.4~eV for the 1~QL
 slab --Figure~\ref{isol} (g)--and decreases down to 0.2~eV for the 6~QL thin film --Figure~\ref{isol} (k)--.
 On the other hand, the 6~QL \bite\ slab under purely biaxial strain
is a TI which develops TSSs with no gap, and its DP 
lies below the Fermi level --Figure~\ref{isol} (a)--. When the two subsystems are brought
 together, the bulk--to--boundary correspondence dictates that
 topologically protected interface states must develop in the gap.
 The band structures of the junctions are shown in Figure~\ref{NI} (a) to (f).

\begin{figure}
 \centering
\includegraphics[scale=0.55]{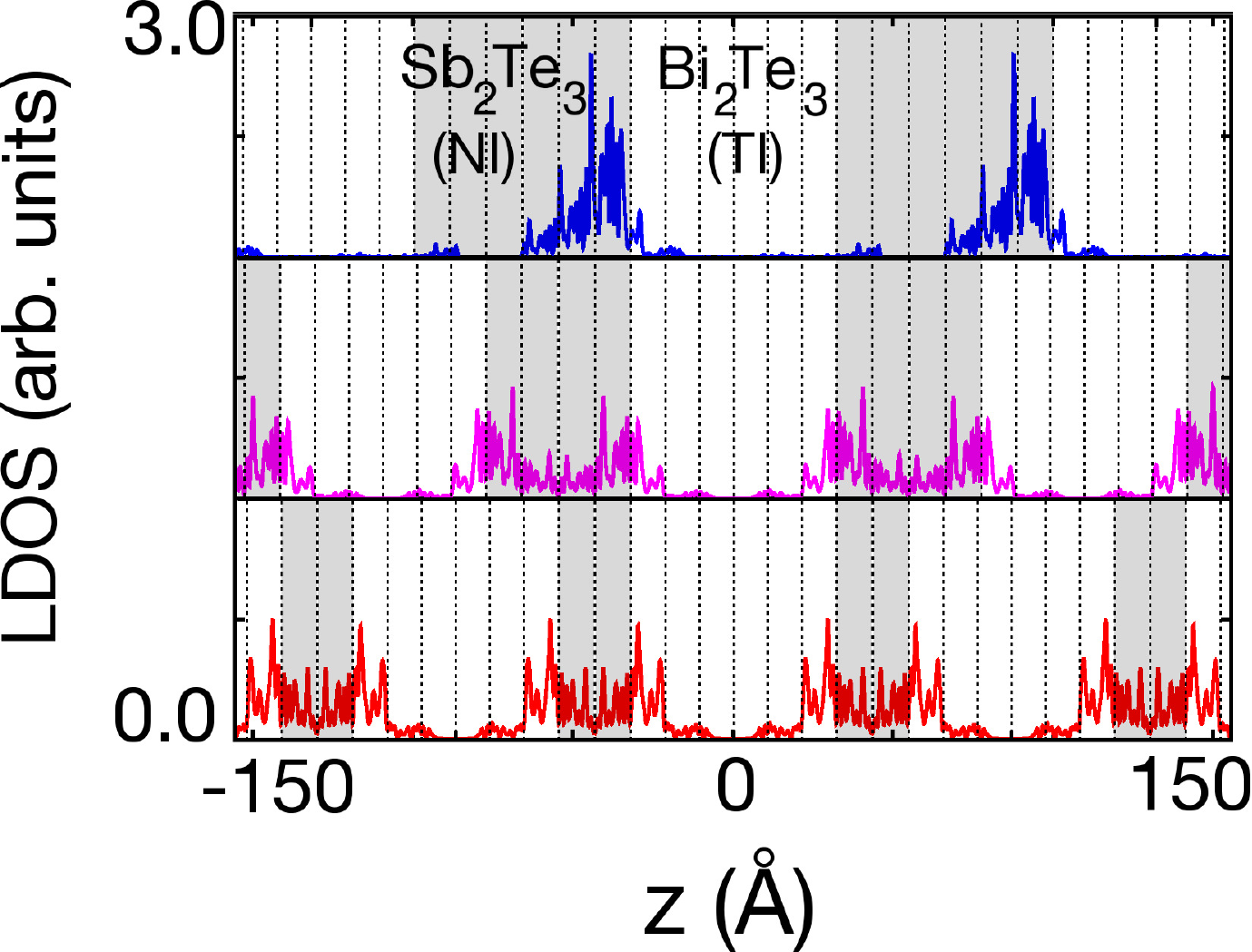}
 \caption{LDOS of the TISs in $m$-\sbte/$n$-\bite/$m$-\sbte\ 
superlattices
for $m$=1 (red line at the bottom), $m$=2 (magenta line in the 
middle) and $m$=3 (blue line at the top) --note that the total thickness of the \sbte\
subsystem is 2$m$ QLs--.
The LDOS was computed at $\bb{k}$=(5, 0) $\cdot 10^{-3}$\AA$^{-1}$, along the $\bar{\Gamma}$--$\bar{K}$ direction
 and close to the $\bar{\Gamma}$ point for the electron--like TISs.
The system is centered at the middle of the 6~QL \bite\ layer. Vertical dashed lines
depict the boundaries of each QL. The gray shaded region corresponds to the \sbte\
 subsystem, while white regions belong to the \bite\ subsystem.
TISs of NI/TI superlattices exhibit strong
 hybridization with the opposite interface for thicknesses of the $2m$--\sbte\ layer below $m$=2 QLs, while
 for the $m$=3 QLs the TISs are already decoupled.
 All three TISs shown are degenerate due to inversion symmetry,
 and only one of the two--fold degenerate states is shown in each case.
}
 \label{NITIpens}
\end{figure}

 For periodic boundary conditions --panels (a) to (c) of Figure~\ref{NI}--, 
and in contrast to the TI/TI superlattices analyzed 
 in subsection~\ref{ssec:TITI}, a topological interface state (TIS) develops that spans the bulk band gap.
 Unexpectedly, the interface topological state localizes in the normal insulator \sbte\ (see Figure~\ref{NITIpens}),  as opposed to TSSs,  
 which  always localize in the topological insulator.
 In this heterojunction a hybridization gap opens in the spectrum for thicknesses of 
 the $2m$--\sbte\ layer below $m$=2 QLs --Figure~\ref{NI} (b)--
 since the two opposed interfaces are closer
 than twice the penetration depth of the TISs. The TISs
 show no doping in contrast with the TSS of 6~QL \bite\ --see Figure~\ref{NI} (f)--, opening 
  a new way of tuning the DP energy of the topological states.

\begin{figure}
 \centering
\includegraphics[scale=0.68]{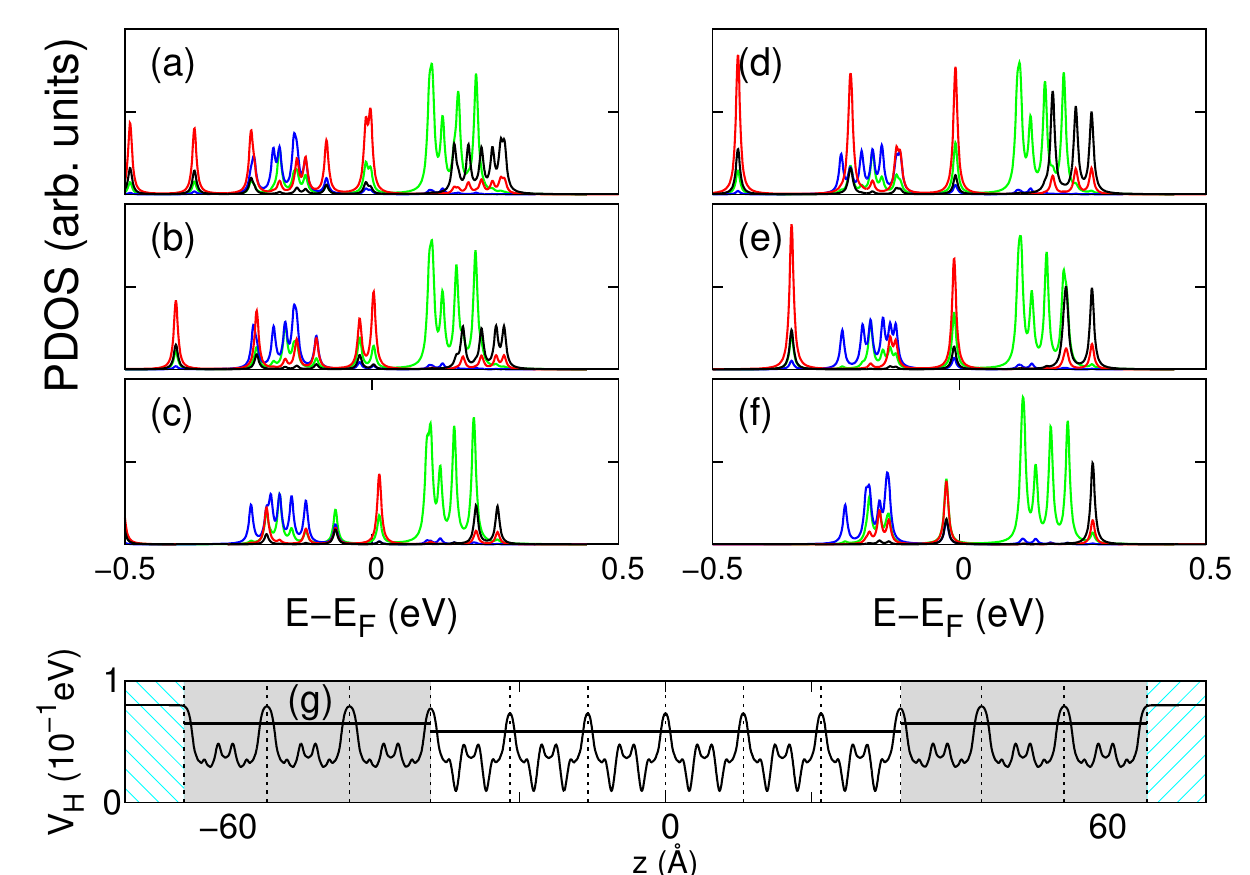}
 \caption{(a) to (f) show the PDOS in $\bar{\Gamma}$ close to the Fermi level of the studied NI/TI heterojunctions.
Blue and black lines show the contribution of Bi $p_z$ and Sb $p_z$ orbitals respectively.
The green and red lines indicate the Te $p_z$ contributions from the Te atoms in the \bite\ and \sbte\ subsystems respectively.
(a) to (c) --(d) to (f)-- depict the PDOS for superlattices --slabs-- with $m$=3 to 1. The \bite\ subsystem shows band inversion, while
 the \sbte\ subsystem is uninverted. The 2D averaged Hartree potential profile
 for the $m$=3 NI/TI slab is shown in (g), along with the average Hartree potential in each QL (horizontal straight solid lines).
}
 \label{dosNITI}
\end{figure}

 To understand the fundamental difference between the TI/TI and NI/TI heterojunctions,
we analyze the atomic orbital PDOS at $\bar{\Gamma}$ in Figs.~\ref{dosTITI} and~\ref{dosNITI}.
 While the 
\bite\ shows band inversion in both cases, the \sbte\ subsystem exhibits opposite 
traits in the two different sets of heterojunctions. In the TI/TI systems, as discussed previously,
 there is band inversion, 
while in the NI/TI there is not. In the NI/TI case the Te $p_z$ orbital of 
the \sbte\ with negative parity occupies the top of the VB region, while the 
Sb $p_z$ orbital 
with positive parity is located at the CB just above the Te derived bands of 
\bite. Therefore, the \sbte\ remains in the trivial state. Nevertheless, 
the topological interface states are mainly formed from the orbitals 
closest to the energy gap, namely from the Te orbitals of \sbte. This feature 
explains why the TISs are located predominantly in the non-topological slab. Furthermore,
due to their spatial localization, there is a strong 
interaction between the TISs at both interfaces for $m$=1 and 2, and a gap opens up.

 For the trilayer slab configurations shown in panels (d) to (f) of Figure~\ref{NI}, the
 system shows a common bulk gap of $\sim$~0.15~eV and a gapless interface state
 with the DP at the Fermi level. This topological interface state is also undoped and
 develops irrespective
 of the thickness of the non-topological \sbte\ layers, and analogously to
the emergent TISs in the SLs, it is not 
 strictly localized at the interface. Instead, the state at the gap
 is confined in the \sbte\ subsystem, with more weight at the interfacemost 
 QL of \sbte, but exceeding the expected $\sim$~2~QL penetration depth of the TSSs in an
 isolated \bite\ slab --see Figure~\ref{slabpens} (b)--. 
 Moreover, the orbital distribution in the VB and CB, and hence that of
the TIS, is similar in the NI/TI superlattices and trilayers (see Fig.~\ref{dosNITI}). 
The main difference between the superlattice and the trilayer TISs lies in the lack of 
interaction in the trilayer geometry due to the localization in the \sbte\ subsystem. Thus, all
the trilayers remain semimetallic. On the other hand,  
only minor differences between the averaged Hartree potential of NI/TI and 
TI/TI heterostructures of equivalent TI/TI geometry --see Fig.~\ref{dosTITI} (g) and Fig.~\ref{dosNITI}-- are observed.

 Our findings are in agreement 
 with previous results~\cite{bisesbse,NITISergei,inverseti,Menschikova2013}
 in which similar TISs with large penetration depths appear in NI/TI
 junctions localized in the NI.
 We additionally checked that in NI/TI/NI heterojunctions
 of $m$-\sbte/$n$-\bite/$m$-\sbte\ with $n$ as low as 1~QL, the TISs are always gapless in slab configuration,
 and remain gapless in superlattice geometries as long as $m > 2$. Note 
  that \bite\ here is the TI, and 1~QL \bite\ thin films show a relatively
  large Dirac gap due to surface--surface hybridization. Therefore, 
  capping \bite\ with uniaxially strained \sbte\ leads to a closing of the Dirac gap in the topological states,
  since the latter localizes in the normal insulating \sbte.
 In addition, although the development of TISs in NI/TI junctions is dictated by topology, their spatial location is
 determined by the orbitals dominating the edges of the valence and conduction band
 of the heterostructures, and thus by the relative alignment of the bands of both subsystems.

\section{Summary and conclusions}
\label{sec:summ}
We have shown the combined effects of uniaxial and biaxial strain on \bite, \sbte, \bise\ 
and \sbse, both in bulk and slab geometries. A phase diagram for the four systems was computed 
and analyzed, demonstrating that topological phase transitions, either to a metal or to a trivial insulator, can occur for 
 different combinations of both kinds of strains, and
 a universal behavior was found for the four compounds.
 We showed how strain can engineer the DP energy, the Fermi velocity, the metallic character
 and the topology of the four compounds, thus offering a wide tunability regarding straintronics.
We have also calculated the electronic structure of \sbte/\bite/\sbte\ trilayers, in which \sbte\
was driven into the topologically trivial insulating regime by applying uniaxial strain.
For the TI/TI systems no trivial nor topological interface state is found, and the superlattice shows
a straddling gap of $\sim$~0.1~eV. In the NI/TI 
heterojunctions, topologically protected interface states are predicted and 
characterized.
 Since the TIS spatial location is determined by the relative band alignment
of the two compounds forming the heterostructures,       
 we find
 TISs to localize in the NI both in slab configurations and periodic superlattices,
 thus opening a route to closing hybridization gaps in topological states
  of ultrathin films of the \bise\ family by capping the system with NI layers. 
Our results for the NI/TI heterojunctions also indicate a way to avoid interactions
 of the topological states with undesired ambient impurities while
 preserving the bulk band gap of the system, and thus maintaining
 the topological protection of the states.
 Uniaxial strain on the \sbte\ subsystem can additionally turn the interface conducting channel 
on or off, thus the system hosts a switchable topological interface state irrespective of the thickness of the TI
layer.

\section*{Acknowledgments}
This work has been supported by the Spanish Ministry of Economy and Competitiveness 
 through Grant MINECO/FEDER No. MAT2015-66888-C3-1R. We acknowledge the use of computational resources of 
 CESGA, Red Espa\~nola de Supercomputaci\'on (RES) and the i2BASQUE academic 
 network.
\bibliography{Biblio.bib}
%\begin{thebibliography}
%\end{thebibliography}

\appendix*
\section{Band dispersion tables for \bite, \sbte\ and \sbse}
 \begin{figure*}
 \includegraphics[scale=0.50]{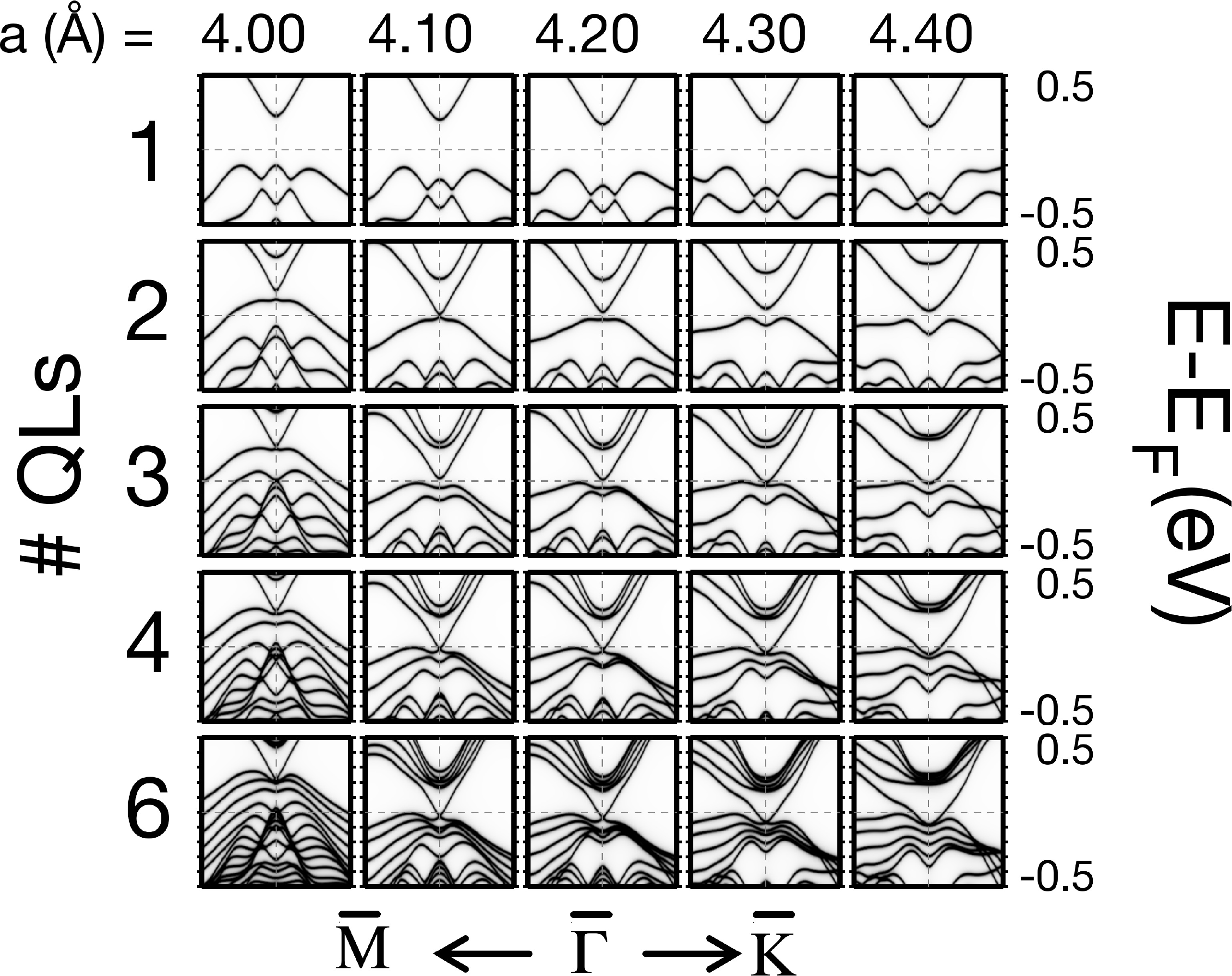}
 \caption{Same as Figure~\ref{bisestress} for \bite.}
 \label{bitestress}
 \end{figure*}

 \begin{figure*}
 \includegraphics[scale=0.50]{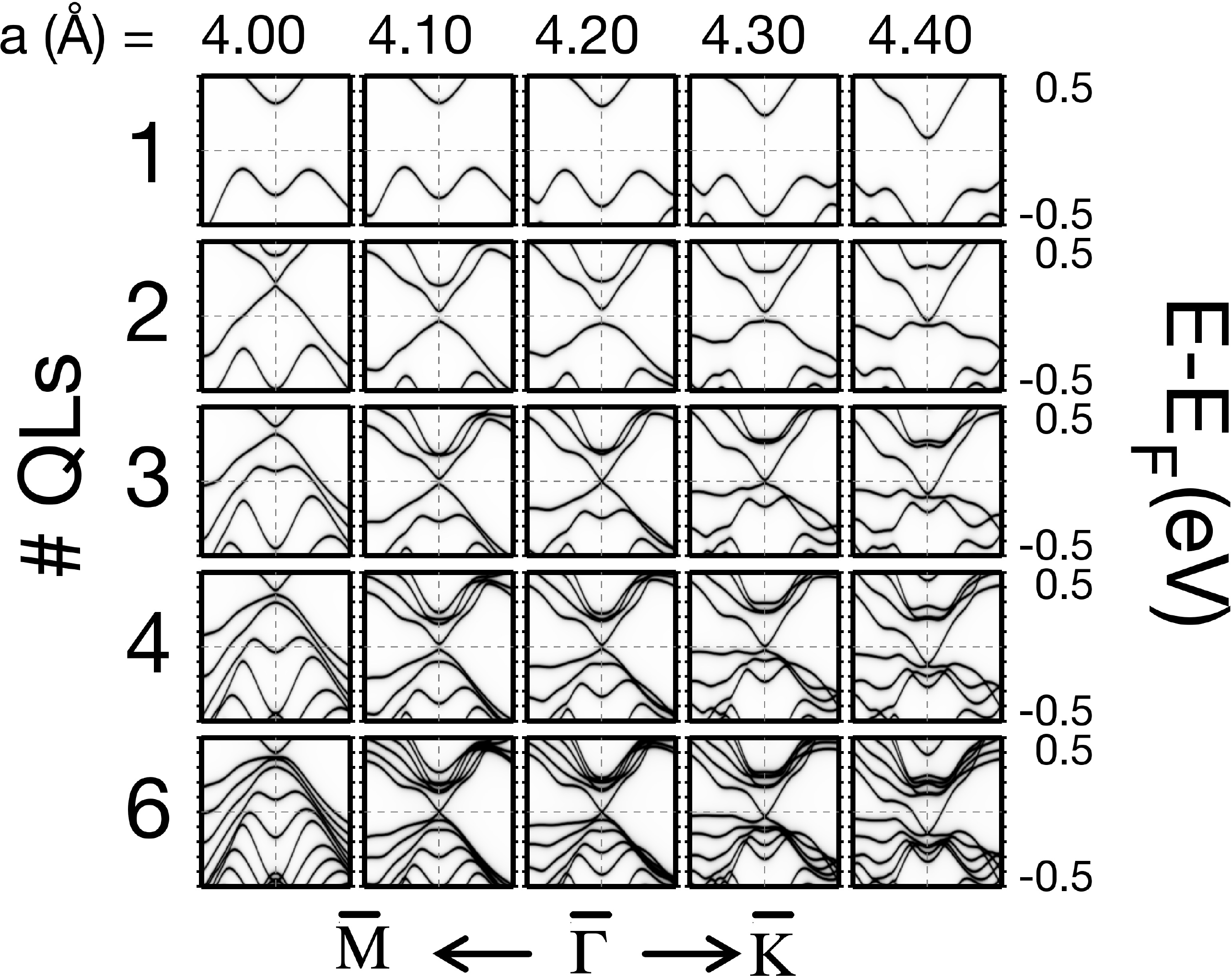}
 \caption{Same as Figure~\ref{bisestress} for \sbte.}
 \label{sbtestress}
 \end{figure*}

 \begin{figure*}
 \includegraphics[scale=0.50]{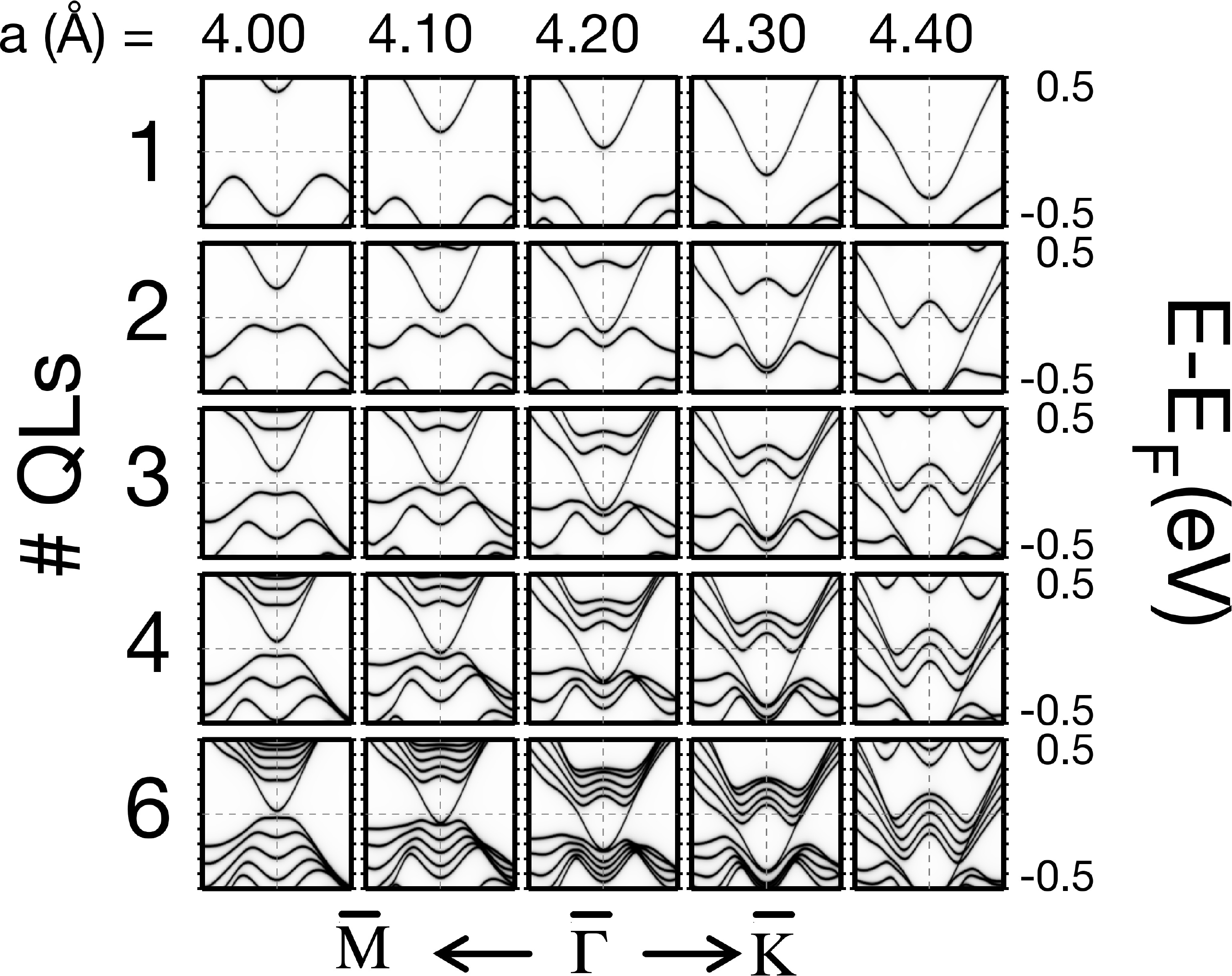}
 \caption{Same as Figure~\ref{bisestress} for \sbse.}
 \label{sbsestress}
 \end{figure*}

\end{document}